\pdfoutput=1

\documentclass[11pt]{article}

\usepackage{ACL2023}

\usepackage{times}
\usepackage{latexsym}

\usepackage[T1]{fontenc}

\usepackage[utf8]{inputenc}

\usepackage{microtype}

\usepackage{inconsolata}
\usepackage{algorithmic}
 \usepackage{array,multirow,graphicx}
\usepackage[ruled,vlined,linesnumbered]{algorithm2e}
\usepackage{comment}
\usepackage{float}
\usepackage{colortbl}
\usepackage{xcolor}
\usepackage{varioref}
\usepackage{pgfplots} 
\usepackage{subcaption}

\title{Subjective Crowd Disagreements for Subjective Data: Uncovering Meaningful \emph{CrowdOpinion} with Population-level Learning}

\author{\begin{tabular}{c} Tharindu Cyril Weerasooriya \textsuperscript{1*},
Sarah Luger\textsuperscript{2},
Saloni Poddar\textsuperscript{1},\\
Ashiqur R. KhudaBukhsh\textsuperscript{1},
Christopher M. Homan\textsuperscript{1}\end{tabular}\\
  \textsuperscript{1}Rochester Institute of Technology, USA\\
   \textsuperscript{2}Orange Silicon Valley\\
   {\textsuperscript{*}\tt cyriltcw@gmail.com}\\
   }
\begin{document}

\maketitle

\begin{abstract}
\textcolor{red}{This paper contains content that can be offensive or disturbing.}

Human-annotated data plays a critical role in the fairness of AI systems, including those that deal with life-altering decisions or moderating human-created web/social media content. Conventionally, annotator disagreements are resolved before any learning takes place. However, researchers are increasingly identifying annotator disagreement as pervasive and meaningful. They also question the performance of a system when annotators disagree. Particularly when minority views are disregarded, especially among groups that may already be underrepresented in the annotator population. In this paper, we introduce \emph{CrowdOpinion}\footnote{Accepted for publication at ACL 2023}, an unsupervised learning based approach that uses language features and label distributions to pool similar items into larger samples of label distributions. We experiment with four generative and one density-based clustering method, applied to five linear combinations of label distributions and features. We use five publicly available benchmark datasets (with varying levels of annotator disagreements) from social media (Twitter, Gab, and Reddit). We also experiment in the wild using a dataset from Facebook, where annotations come from the platform itself by users reacting to posts. We evaluate \emph{CrowdOpinion} as a label distribution prediction task using KL-divergence and a single-label problem using accuracy measures. 
\end{abstract}

\section{Introduction}

\begin{figure} [t]

\begin{subfigure}[t]{0.15\textwidth}
\begin{center}
\includegraphics[width=1\textwidth]{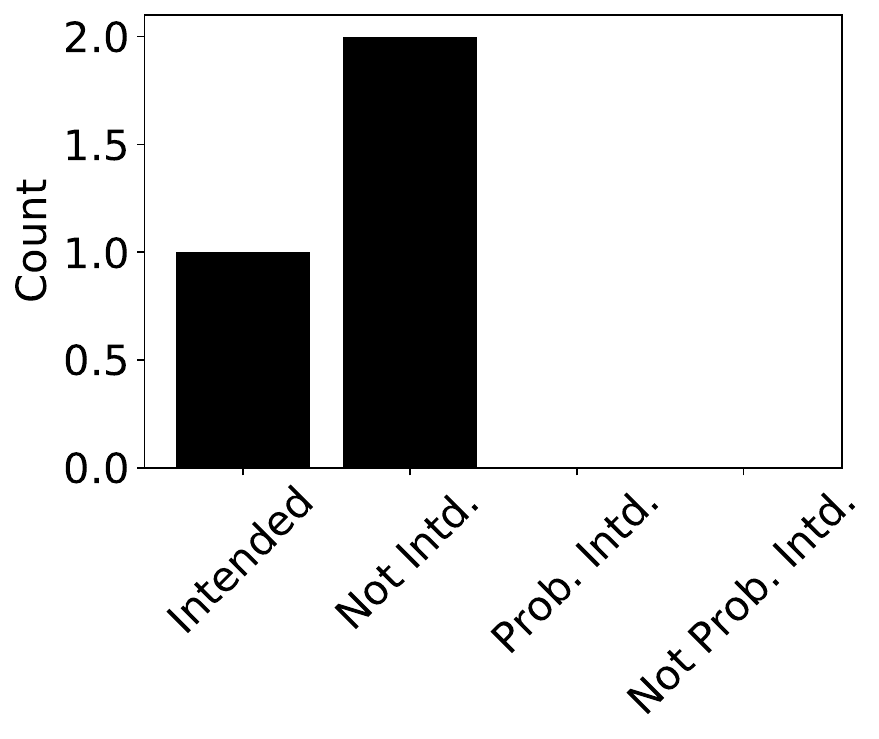}
    \caption{$\mathcal{D}_\texttt{SI}$E1 ``During the Thanksgiving season, many Americans support Jennie-O-cide.''}
    \label{fig:SI_example1}
\end{center}
\end{subfigure}\hfill
\begin{subfigure}[t]{0.15\textwidth}
\begin{center}
\includegraphics[width=1\textwidth]{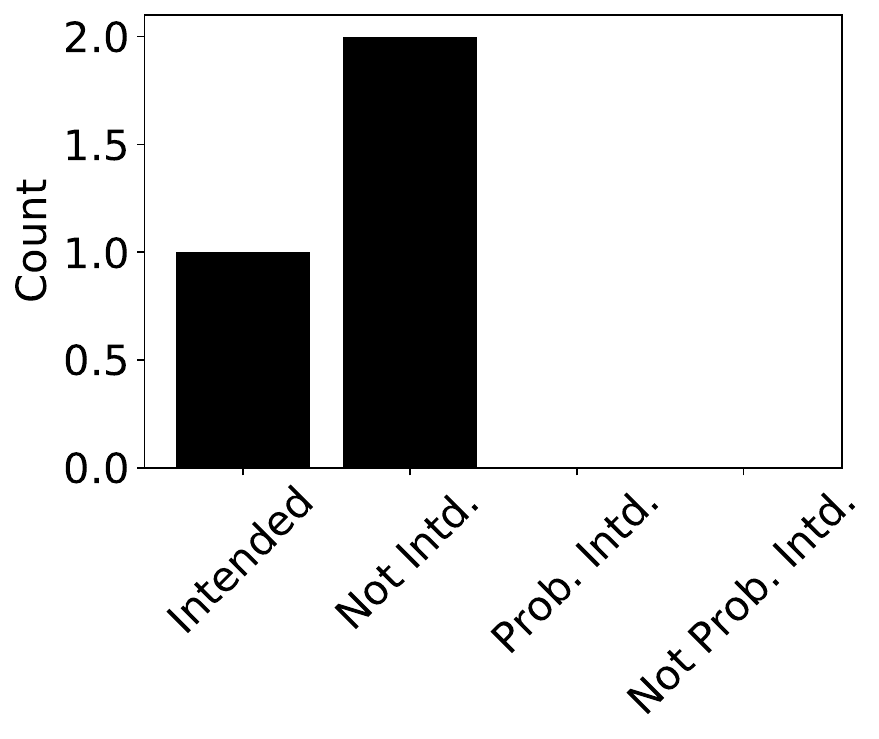}
    \caption{$\mathcal{D}_\texttt{SI}$E2 ``That's not your real name. You're supposed to have a foreign name or something.''}
    \label{fig:SI_example2}
\end{center}
 \end{subfigure}
 \hfill
\begin{subfigure}[t]{0.15\textwidth}
\begin{center}
\includegraphics[width=1\textwidth]{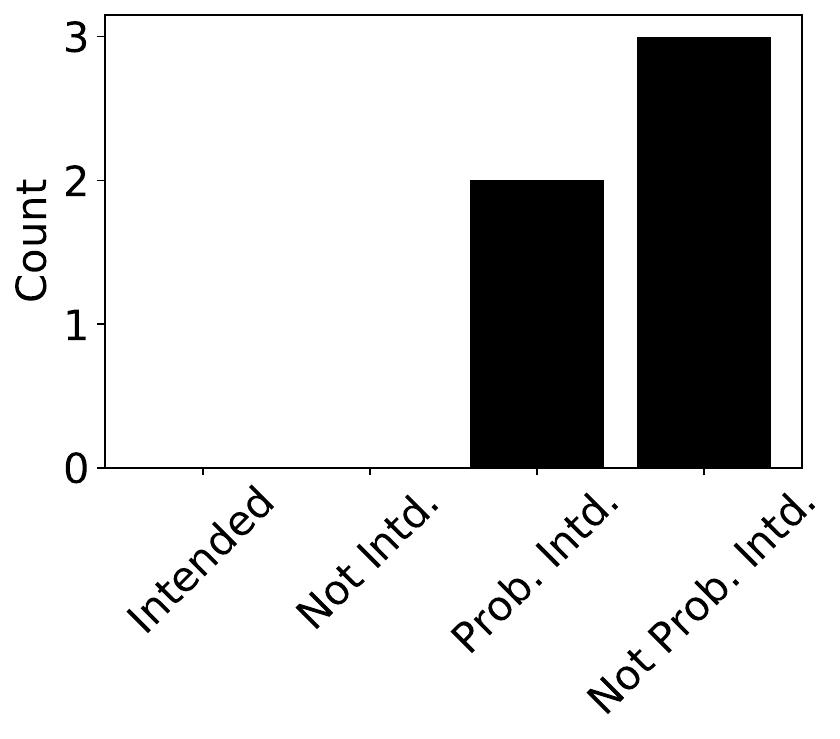}

    \caption{$\mathcal{D}_\texttt{SI}$E3 ``what's the difference between jelly and jam?  i can't jelly my d*** down your throat.''}
     \label{fig:SI_example3}
\end{center}
 \end{subfigure}
    \caption{Examples from $\mathcal{D}_{SI}$ \cite{sap2019social}, from human annotation for Twitter posts on whether they are intended to be offensive. These examples show how offense cannot generalize, and in cases when a majority of the annotators are not offended the input for a classifier is the majority voice.}
 \label{fig:SI_examples}
\vspace{-0.5cm}
\end{figure}

Long term exposure to offensive, threatening, and hate speech posts through any public-facing social media platform can lead to depression or even physical injuries, specially at a younger age \cite{pedalino_instagram_2022}. This is a persistent problem in social and web content where the impact could be not limited to just the targeted parties but expand to anyone in the community consuming the content \cite{benson1996rhetoric,fauman2008cyber,chandrasekharan2017you,HateCrimeCausal}. 

Language used by content creators in social media (see Figure \ref{fig:SI_examples}) with a subtle tone and syntax can hide the offensive content from the purview \cite{basile2019semeval,zubiaga_detection_2019} or machine learning classifiers \cite{kumar_designing_2021}. This challenge has ethical and legal implications in many countries as these governments have imposed restrictions for platforms to identify and remove such harming content \cite{kralj_novak_handling_2022,saha_prevalence_2019} citing the right for safety. 

\begin{figure} [t]
    \centering
    \includegraphics[width=.45\textwidth]{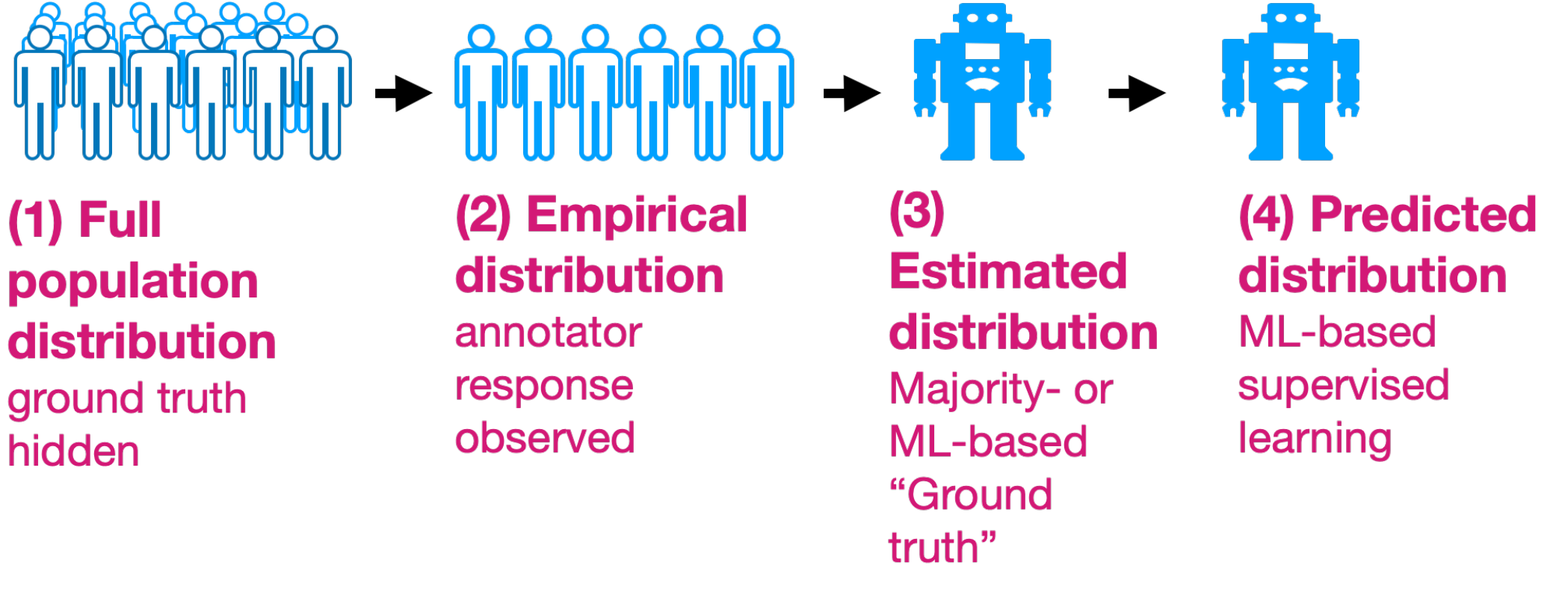}
    \caption{\small{Each learning example is associated with sets of labels: (1) the (hidden) full distribution of responses by the entire population of annotators/stakeholders; 
    (2) the (observed) responses received often from human crowdworkers hired to annotate the data; 
    (3) an estimate of the full population distribution based on the empirical distribution of the given item (and frequently of other items and anonymized identifiers for the annotators); 
    (4) the prediction of a machine learning model trained on either the empirical or estimated distributions.}}
    \label{fig:layers}
\vspace{-0.5cm}
\end{figure}

The ML classifiers generally rely on human feedback \cite{Eriksson2010,Dong2019}. Because humans, as content creators or annotators (content moderators), are subjective in their opinions \cite{alm_subjective_2011}. Their feedback is essential to understanding subjective web or social media content. The standard practice is to ask multiple annotators about each post and then use the majority opinion or ML-based methods to determine the ground truth label (see Figure \ref{fig:layers}). 

Typically, minority views are completely removed from the dataset before it is published. Yet these views are often meaningful and important \cite{aroyo2014,kairam2016parting,plank2014linguistically,chung2019efficient,Obermeyer2019,founta2018large}. Figure \ref{fig:SI_examples} shows three tweets with offensive language that have been labeled by multiple annotators about the tweeter's intent \cite{sap2019social}. In each case, the majority of annotators considers the offensiveness to be \emph{not intended}. Yet a minority considers it to be \emph{intended}. 
A classifier trained on such language data after these minority opinions are removed would not know about them. This is dangerous because abusers often obscure offensive language to sound unintended in case they are confronted \cite{smits_origin_2022}. And so, removing minority opinions could have dramatic impacts on the model's performance if, say, it was trying to detect users creating hateful or offensive content on a social platform. 

Consequently, a growing body of research advocates that published datasets include ALL annotations obtained for each item \cite{Geng2016,Liu2019HCOMP,Klenner2020,Basile2020,prabhakaran-etal-2021-releasing}. And a substantial body of research is studying annotator disagreement \cite{aroyo2014,kairam2016parting,plank2014linguistically,chung2019efficient,Obermeyer2019,founta2018large,binns2017}.
Unfortunately, most existing datasets are based on 3--10 annotators per label, far too few, statistically speaking, to represent a population. Thus, learning over such a sparse space is challenging.

\citet{Liu2019HCOMP} show that clustering in the space of label distributions can ameliorate the sparseness problem, indicating that data items with similar label distributions likely have similar interpretations. Thus, a model can pool labels into a single collection that is large enough to represent the underlying annotator population. Recent work by \citet{davani_dealing_2022}, studying annotator disagreement with majority vote and multi-label learning methods, has called out the need for cluster-based modeling to understand annotator disagreements.

The lack of annotator-level labels also hinders studying the annotator behaviors using methods that utilize those granular-level labels \cite{dawid1979,rodrigues2018deep,gordonJuryLearningIntegrating2022,collins_eliciting_2022,liu2023second}. We see this as a benefit to \emph{CrowdOpinion} (CO) we propose, a technique applicable at a broader level for understanding and predicting annotator disagreements which mitigate granular-level annotations. 

The \textbf{motivation} behind \emph{CrowdOpinion} is to reduce inequity and bias in human-supervised machine learning by preserving the full distribution of crowd responses (and their opinions) through the entire learning pipeline. We focus our methods on web and social media content due to its subjectivity. Our contributions to this core problem in AI and NLP is a learning framework\footnote{Experimental code available through \url{https://github.com/Homan-lab/crowdopinion}} that uses unsupervised learning in Stage 1 on both the labels {\bf AND} data features to better estimate soft label distributions. And in Stage 2, we use these labels from Stage 1 to train and evaluate with a supervised learning model. We consider the following three questions.

\textbf{Q1}\label{Q_method}: \emph{Does mixing language features and labels lead to better ground truth estimates than those that use labels only?} This focuses on the first stage as a standalone problem and is difficult to answer directly, as ``ground truth'' from our perspective is the \emph{distribution of labels from a hidden population of would-be annotators}, of which we often only have a small sample (3-10 annotators) per data item. 
We study four generative and one distance-based clustering methods, trained jointly on features and label distributions, where we vary the amount of weight given to features versus labels.

\textbf{Q2}\label{Q2}: \emph{Does mixing features and labels in the first stage lead to better label distribution learning in the second?} 
We use the label distributions obtained from the first-stage models from \textbf{Q1} as feedback for supervised learning.
 We compare our results with baselines from pooling based on labels only \cite{Liu2019HCOMP}, predictions trained on the majority label for each item without clustering, and predictions trained on the label distribution for each item but without any other first-stage modeling. Our results show improvement over unaggregated baselines. 

\textbf{Q3}\label{Q_sl}: \emph{Do our methods lead to better single-label learning (SL)?} Since most applications consider only single-label prediction, we measure the model performance on single-label prediction via accuracy. 

\subsection{Beyond Experiments} 
Humans have annotated our benchmark datasets for specific tasks. However, this is not always the case in practice. Social networks have introduced \emph{reactions} that allow users to react to platform content. We study this use case by predicting these reactions for Facebook posts \cite{WolfFb} as a special case.
 

\begin{figure}[t]

\tiny
\centering

\tiny
\centering
\begin{subfigure}[t]{0.15\textwidth}
\begin{center}
\includegraphics[width=1\textwidth]{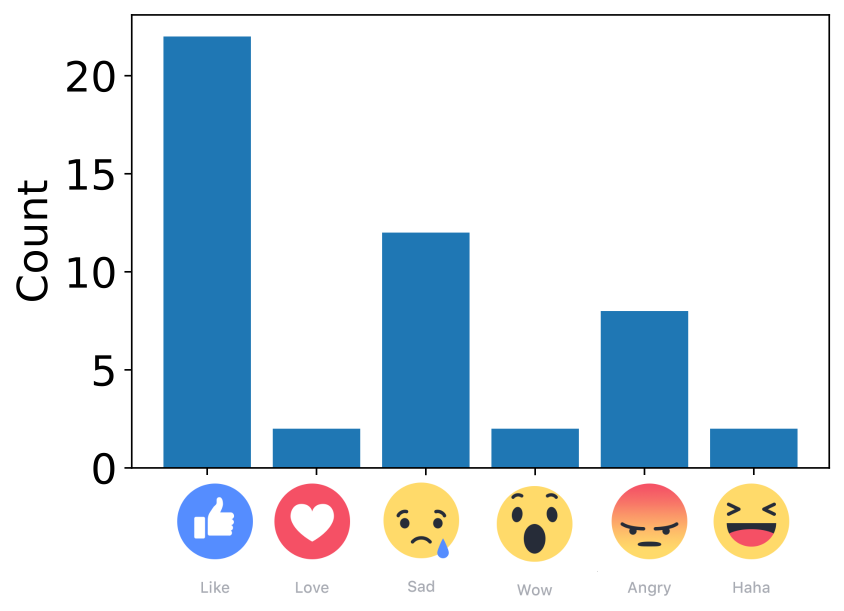}
    \caption{$\mathcal{D}_\texttt{FB}$E1 ``Some are calling President Obama the ``deporter-in-chief,'' rewriting his legacy.''}
    \label{fig:fb_example1}
\end{center}
\end{subfigure}\hfill
\begin{subfigure}[t]{0.15\textwidth}
\begin{center}
\includegraphics[width=1\textwidth]{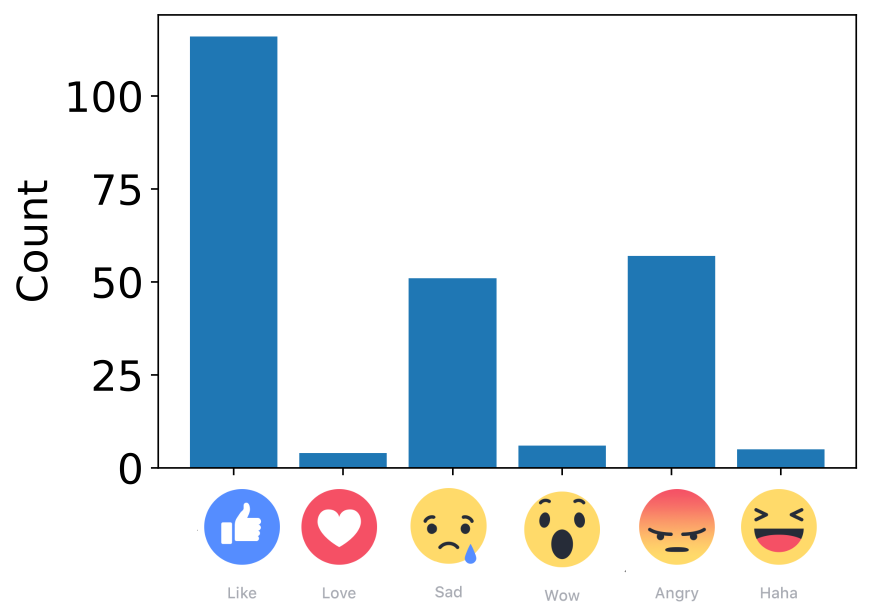}
    \caption{$\mathcal{D}_\texttt{FB}$E2 ``Some criticized Obama administration attempting to deport immigrant women and children.''}
    \label{fig:fb_example2}
\end{center}
 \end{subfigure}
 \hfill
\begin{subfigure}[t]{0.15\textwidth}
\begin{center}
\includegraphics[width=1\textwidth]{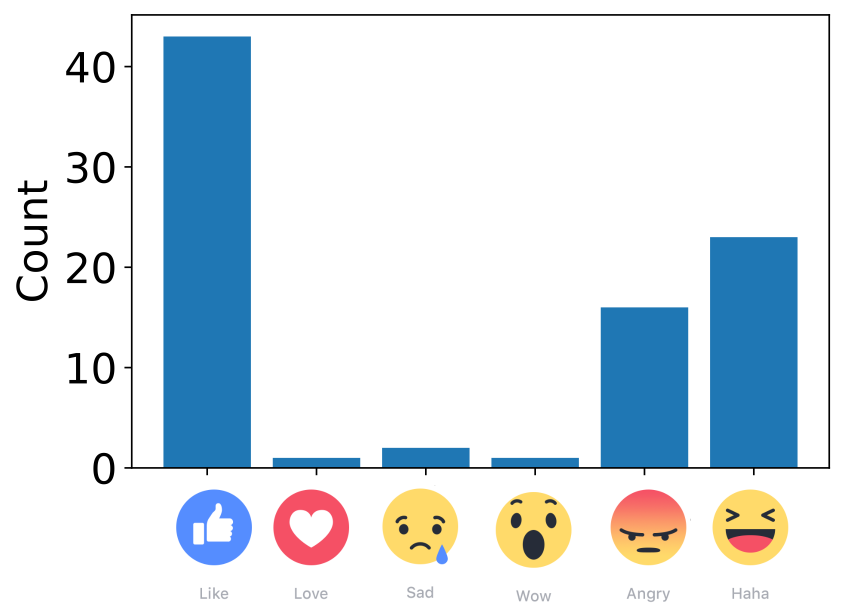}
    \caption{$\mathcal{D}_\texttt{FB}$E3 ``Ted Cruz expressed that President Obama, instead of standing with our allies, insults them continuously.''}
     \label{fig:fb_example3}
\end{center}
 \end{subfigure}
    \caption{\small{Motivating examples from $\mathcal{D}_{FB}$ \cite{WolfFb}, demonstrating human disagreement to three posts. Reactions are \textit{like, love, sad, wow, angry, and haha.}}}
 \label{fig:fb_examples}
\vspace{-0.5cm}
 \end{figure}
 
Among the top 100 posts from Facebook (entropy $>$ 1.2), 26 were about Donald Trump, with most of the label distribution mass divided between ``like'', ``haha'', and ``angry''. Another 26 posts were about politics (but not Trump), with the label distribution mass generally divided between ``angry'' and ``sad''. There were only two non-English posts and no sports-related posts. And interestingly, except for two non-English posts, all of the other top posts had a substantial portion of their mass on ``angry''.

The bottom 100 set (entropy $<0.04$) contains 46 posts about sports and 13 non-English posts. There was only one political post (and it was not about Trump). The label distribution pattern in this set was more dominated by ``like'' ($>98\%$), followed by reactions of either ``love'' or ``haha''. ``Like'' was also dominant in the high entropy posts, but not to such a degree; based on this observation and \cite{tian2017facebook}, we eliminate it from our experiments. 

Figure~\ref{fig:fb_examples} illustrates some nuances in meaning that different label distributions reveal. All three are negative posts about Barack Obama, and all have most of their mass on ``like''. $\mathcal{D}_\texttt{FB}$E1 and $\mathcal{D}_\texttt{FB}$E2 have similar distributions, in contrast to $\mathcal{D}_\texttt{FB}$E3 where, besides ``like'', the distribution mass falls mainly on ``haha'' and ``angry''.  Perhaps this is because, in contrast to the first two posts which are from anonymous sources, the criticism on $\mathcal{D}_\texttt{FB}$E3 comes from a political rival, and maybe this provides a concrete target for ridicule? 

\subsection{Facebook's Special Case}
"Like" was the original Facebook reaction and platform users may find it a quick, default, and intuitive interaction. The over-representation of "like" on Facebook exemplifies how this dataset is an unusual human annotation case. It is unique not only in the human labeling behavior, but also in the resulting label distribution. 
%

\section{Methods - \emph{CrowdOpinion}}
\label{sec:methods}

In conventional, nondistributional supervised learning, clustering might happen over the feature space only as a form of data regularization \citep{nikulin2009regularised}; the labels, being strictly categorical and nondistributional, would be scalar and thus too simple to benefit from extensive modeling. In our setting, each data item $x_i \in \mathcal{X}$ is associated with a vector $y_i \in \mathcal{Y}$, representing the empirical distribution of ALL annotator responses, which we view as \emph{sample} of a larger, hidden population. Our approach, \emph{CrowdOpinion} (CO) is two-staged and summarized in Algorithm \ref{alg:clpldl}. 
 
 In Stage 1, we cluster together related data items and share among them a label distribution $\hat{y}_i$ based on all labels from all items in each cluster. This stage resembles, in function, a deep vein of label estimation research begun by Dawid and Skene \cite{dawid1979,carpenter2008multilevel,ipeirotis2010quality,pasternack2010knowing,weld2011human,raykar2012eliminating,kairam2016parting,Mitchell2021}, except that (a) our output is an estimate of the distribution of label responses by the underlying population of annotators, not a single label, and (b) $y_i$ in their models is a vector with one dimension for each annotator. To better handle the label sparseness common in most datasets, our $y_i$ has one dimension for each label choice, representing the proportion of annotators who made that choice. Stage 2 performs supervised learning on these new item, label distribution pairs $(x_i, \hat{y}_i)$.


\begin{algorithm}[h]

\SetAlgoLined
\caption{\small CO-$\mathcal{C}$-$\mathcal{H}$-$w$}
\label{alg:clpldl}
\SetKwProg{generate}{Function }{}{end}
\textit{\textbf{Parameters:}}\\
Clustering (or pooling) algorithm $\mathcal{C}$\\
Hypothesis space $\mathcal{H}$\\
Mixing parameter $w \in [0,1]$\\
\textit{\textbf{Inputs:}}\\
Data features with empirical label distributions $(x_i, y_i)_{1\leq i \leq n}$  \hspace{1cm} // BOTH $x_i$ and $y_i$ are vectors!\\
\textit{\textbf{Procedure:}}\\
Stage 1:\\
~~Perform clustering with $\mathcal{C}$ on BOTH item features and labels, weighted and concatenated together: $(w \cdot x_i, (1-w) \cdot y_i)_{1 \leq i \leq n}$\\
~~Let $(\hat{x}_i, \hat{y}_i)$ be the centroid of the cluster $\pi_j$ associated with each $(x_i, y_i)$\\
\label{cluster}
Stage 2: 
~~Perform supervised learning on $(x_i, \hat{y}_i)$ over hypothesis space $\mathcal{H}$\\ \label{learn}
\end{algorithm}
Note that nearly any pair of clustering $\mathcal{C}$ and supervised learning $\mathcal{H}$ algorithms can be used for stages one and two, respectively. \citet{Liu2019HCOMP} performed the same kind of label regularization only using the label space $\mathcal{Y}$, it is a baseline for our methods ($w=0$). Our main technical innovation is to perform label regularization based on the \emph{weighted joint feature and label} space $w \cdot \mathcal{X} \times (1-w) \cdot \mathcal{Y}$, where $w \in [0,1]$ is the \emph{mixing parameter} that determines the relative importance of $\mathcal{X}$ versus $\mathcal{Y}$ during clustering. 



 We consider four clustering models $\mathcal{C}$ 
 used by \citet{Liu2019HCOMP}: a (finite) multinomial mixture model (\textbf{FMM}) with a Dirichlet prior over $\pi \sim \mbox{Dir}(p, \gamma = 75)$, where $p$ is the number of clusters and each cluster distribution $\pi_j$ is a multinomial distribution with Dirichlet priors $\mbox{Dir}(d, \gamma = 0.1)$, where $d$ is the size of the label space, using the bnpy library \cite{Hughes2013}, a Gaussian mixture model (\textbf{GMM}) and a K-means model (\textbf{KM}) from scikit-learn, and the Gensim implementation of Latent Dirichlet Allocation (\textbf{LDA}) \cite{gensim}. Each of these models takes as a hyperparameter the number of clusters $p$.

We perform parameter search ($4\leq p \leq 40$) on the number of clusters, choosing $\arg\min_p \sum_i KL((x_i, y_i)_w, (\hat{x}_i, \hat{y}_i)_w)$, i.e., the $p$ that minimizes
the total KL divergence between the raw and clustered label distribution, where, e.g.,
$(x_i, y_i)_w$ denotes $(w \cdot x_i, (1-w) \cdot y_i)$, i.e., the weighted concatenation of $x_i$ and $y_i$.

We also consider a soft, distance-based clustering method, called \emph{neighborhood-based pooling} (\textbf{NBP}) in the context of PLL \cite{Weerasooriya2020}. For each data item $i$ it averages over all data items $j$ within a fixed Kullback-Liebler (KL) ball of radius $r$:
\begin{equation}
    \nonumber \hat{y}_i  =  \overline{\{y_j~|~KL\left((x_i, y_i)_w \| (x_j,y_j)_w\right) < r\}}.
\end{equation}

Here, the hyperparameter is the diameter $r$ of the balls, rather than the number of clusters, and there is one ball for each data item. We perform hyperparameter search ($0\leq r \leq 15$) via methods used in \cite{Weerasooriya2020}. Table~\ref{tab:kl_model_summary} summarizes model selection results using these methods.


The supervised model (\textbf{CNN}) for $\mathcal{H}$ is a 1D convolutional neural network \cite{kim2014convolutional}, with three convolution/max pool layers (of dimension 128) followed by a dropout (0.5) and softmax layer implemented with TensorFlow. The input to the model is a 384-dimension-vector text embedding, described below. Table \ref{table:KL_results} summarizes the supervised-learning based classification results. 


We compare our methods against four baselines. \textbf{PD} is our \textbf{CNN} model but with no clustering; it is trained directly on the raw empirical label distributions $(y_i)$. \textbf{SL} the same model, but trained on one-hot encodings of most frequent label in each $y_i$. \textbf{DS+CNN} uses the \citet{dawid1979} model for $\mathcal{C}$ and $\mathcal{H} = $ \textbf{CNN}. \textbf{CO-$\mathcal{C}$-CNN-$0$} is from \citet{Liu2019HCOMP}, which clusters on labels only. 



We represent language features for both our unsupervised learning and classification experiments using a state-of-the-art pre-trained \texttt{paraphrase-MiniLM-L6-v2} transformer model using SBERT (sentence-transformers) library \cite{reimers-2019-sentence-bert}. We identified this pre-trained model based on STS benchmark scores at the time of writing. The feature vector size for each post is 384.

\section{Experiments}
\subsection{Dataset Descriptions}
\label{sec:data}

\begin{table}[h]
\resizebox{\columnwidth}{!}{
\begin{tabular}{ccccc}

    \textbf{Dataset} &\textbf{No. of ants.} &\textbf{Total data} &\textbf{No. of label} &\textbf{Avg.} \\
    &\textbf{(per item)} &\textbf{items} &\textbf{choices} &\textbf{Entropy} \\     \hline
    $\mathcal{D}_\texttt{FB}$ (Facebook) &Avg. 862.3 & 8000 &5 &0.784 \\
    $\mathcal{D}_\texttt{GE}$ (Reddit) &Avg. 4 &54263 &28 &0.866 \\
    $\mathcal{D}_\texttt{JQ1}$ (Twitter) &10 &2000 &5 &0.746 \\
    $\mathcal{D}_\texttt{JQ2}$ (Twitter) &10 &2000 &5 &0.586 \\
    $\mathcal{D}_\texttt{JQ3}$ (Twitter) &10 &2000 &12 &0.993 \\
    $\mathcal{D}_\texttt{SI}$ (Reddit) &Avg. 3 &45318 &4 &0.343 \\
    \hline
    \end{tabular}
    }
    
    \caption{\small{Experimental datasets summary: We calculated entropy per data item and averaged it over the dataset to measure uncertainty. $\mathcal{D}_\texttt{FB}$ \cite{WolfFb}, $\mathcal{D}_\texttt{GE}$ \cite{demszky2020goemotions},  $\mathcal{D}_\texttt{JQ1-3}$ \cite{P16-1099}}, and $\mathcal{D}_\texttt{SI}$ \cite{sap2019social}.}
    \label{tab:dataset}
    
\vspace{-0.2cm}
\end{table}

As our approach focuses on human disagreement, we identified datasets that contain multiple annotators and multiple label choices per data item. We conducted our experiments on publicly available human-annotated English language datasets generated from social media sites (Facebook, Twitter, and Reddit). Each dataset consists of 2,000 posts and employs a 50/25/25 percent for train/dev/test split. Larger datasets are downsampled with random selection to 2,000 for a fairer comparison between them. The datasets vary in content, number of annotators per item, number of annotator choices, and source of content. More detailed descriptions of the datasets are included in the Appendix. 

\subsection{Results}

\begin{table}
    \centering
    \tiny
    \begin{tabular}{c|ccccccc}
    
    \textbf{Dataset} &\textbf{$\mathcal{D}_\texttt{FB}$} &\textbf{$\mathcal{D}_\texttt{GE}$} &\textbf{$\mathcal{D}_\texttt{JQ1}$} &\textbf{$\mathcal{D}_\texttt{JQ2}$} &\textbf{$\mathcal{D}_\texttt{JQ3}$} &\textbf{$\mathcal{D}_\texttt{SI}$} \\
    \hline
    Model & NBP & NBP & NBP & NBP & NBP & K-Means \\
    \hline
    \textbf{KL ($\downarrow$)} &0.070 &0.020 &0.123 &0.133 &0.023 &0.050 \\
    \textbf{$r$}/\textbf{$p$} &3 &0.8 &5.6 &2.8 &10.2 & 35 \\
    \textbf{$w$} &$0.5$ &$0$ &$0.25$ &$0.75$ &$0$ &$1.0$ \\
    \hline
    \end{tabular}
    \caption{\small{Optimal label aggregation model summary with the parameters and KL-divergence. Here $r/p$ is the number of clusters for the generative models and $r$ is the neighborhood size for distance-based clustering.  K-Means is the optimum model for $\mathcal{D}_{SI}$, while NBP (distance-based clustering) is the optimal model for the remaining five datasets. 
    }}
    \label{tab:kl_model_summary}
    \vspace{-0.5cm}
\end{table}

\begin{table*}[!h]
\smaller
\centering
\setlength{\tabcolsep}{3pt}
\begin{tabular}{c|l|ccccccc}
\multicolumn{8}{c}{\textbf{KL-Divergence ($\downarrow$)}}\\
& \textbf{Dataset} &\textbf{$\mathcal{D}_\texttt{FB}$} 
 &\textbf{$\mathcal{D}_\texttt{GE}$} &\textbf{$\mathcal{D}_\texttt{JQ1}$} &\textbf{$\mathcal{D}_\texttt{JQ2}$} &\textbf{$\mathcal{D}_\texttt{JQ3}$} &\textbf{$\mathcal{D}_\texttt{SI}$} \\
\hline
 \parbox[t]{2mm}{\multirow{4}{*}{\rotatebox[origin=c]{90}{Baselines}}}  & \textbf{PD} &0.857$\pm$0.006 &2.011$\pm$0.001 &1.092$\pm$0.004 &1.088$\pm$0.003 &1.462$\pm$0.00 &0.889 $\pm$0.00 \\
& \textbf{DS+CNN} &- &3.247$\pm$0.012 &1.042$\pm$0.005 &1.035$\pm$0.003 &3.197$\pm$0.034 &1.514$\pm$0.067 \\ \cline{2-8}
& Model ($\mathcal{C}$) &GMM &LDA &GMM &K-Means &LDA &FMM \\
& \textbf{KL}, \textbf{$w=0$} &0.684$\pm$0.001 &\textbf{1.987$\pm$0.001} &\textbf{0.427$\pm$0.01} &0.510$\pm$0.001 &\textbf{0.823$\pm$0.001} &\textbf{0.860$\pm$0.026} \\ \hline
& \textbf{$w=$} & $0.75$ &$0.50$ &$1.0$ &$0.25$ &$1.0$ &$1.0$ \\
& \textbf{KL} &\textbf{0.680$\pm$0.001} &1.995$\pm$0.001 &0.450$\pm$0.001 &\textbf{0.499$\pm$0.001} &0.884$\pm$0.001 &0.991$\pm$0.003 \\
\hline
\end{tabular}

 \caption{\small{KL-divergence($\downarrow$) results for the CO-$\mathcal{C}$-CNN-$w$ models from Algorithm \ref{alg:clpldl}, using various choices for clustering $\mathcal{C}$ and feature-label mixing $w$. Here $w=0$ is the baseline from \citet{Liu2019HCOMP,Weerasooriya2020} that uses label distributions in the clustering stage, and $w=1$ means that only data feature are used.  The \textit{best} score is included in the table. Full set of results included in Appendix Table~\ref{table:KL_results_full}.}The \textit{best} score for each dataset bolded.}\label{table:KL_results}
 
\begin{tabular}{c|l|ccccccc}
\multicolumn{8}{c}{\textbf{Accuracy ($\uparrow$)}}\\
& \textbf{Dataset} &\textbf{$\mathcal{D}_\texttt{FB}$} &\textbf{$\mathcal{D}_\texttt{GE}$} &\textbf{$\mathcal{D}_\texttt{JQ1}$} &\textbf{$\mathcal{D}_\texttt{JQ2}$} &\textbf{$\mathcal{D}_\texttt{JQ3}$} &\textbf{$\mathcal{D}_\texttt{SI}$} \\ \hline
\parbox[t]{3mm}{\multirow{7}{*}{\rotatebox[origin=c]{90}{Baselines}}} & \textbf{Others} &- &0.652 &0.82 &0.76 &0.81 &- \\
& \textbf{DS+CNN} &- &0.168$\pm$0.003 &0.684$\pm$0.004 &0.658$\pm$0.003 &0.061$\pm$0.031 &0.508$\pm$0.067 \\
& \textbf{PD} &0.780$\pm0.001$ & \textbf{0.987$\pm$0.001} &0.601$\pm$0.001 &0.800$\pm$0.001 &0.880$\pm$0.020 &0.734$\pm$0.001 \\
& \textbf{SL} &0.790$\pm$0.005 &0.942$\pm0.003$ &0.701$\pm$0.002 &0.810 $\pm$0.001 &\textbf{0.888$\pm$0.030} &0.759$\pm$0.002 \\ \cline{2-8}
& Model ($\mathcal{C}$) &GMM &LDA &GMM &NBP &LDA &LDA \\
& \textbf{Acc. ($\uparrow$)},\textbf{$w=0$} &0.785$\pm$0.001 &0.949$\pm$0.001 &0.891$\pm$0.01 &0.873$\pm$0.001 &0.880$\pm$0.001 &\textbf{0.932$\pm$0.001} \\ \hline
& \textbf{$w=$} &$1.0$ &$1.0$ &$0.75$ &$0.25$ &$0.75$ &$0.5$ \\
& \textbf{Acc. ($\uparrow$)} &\textbf{0.798$\pm$0.001} &0.950$\pm$0.001 & \textbf{0.901$\pm$0.01} &\textbf{0.897$\pm$0.001} &0.883$\pm$0.001 &0.920$\pm$0.045 \\
\hline
\end{tabular}
 \caption{\small{Accuracy($\uparrow$) results for the CO-$\mathcal{C}$-CNN-$w$ models from Algorithm \ref{alg:clpldl}, using various choices for clustering $\mathcal{C}$ and feature-label mixing $w$. Here $w=0$ is the baseline from \citet{Liu2019HCOMP} that uses label distributions in the clustering stage, and $w=1$ means that only data feature are used. Since accuracy is a non-distributional statistic, we use the most frequent label for inference (though not during training; we use the same trained models as in Table \ref{tab:kl_model_summary}). Baselines; \textbf{PD} is trained on empirical distributions, and \textbf{SL} classifier is trained on the most frequent label. \textbf{DS} uses \citet{dawid1979} for label aggregation and our CNN model for prediction. The result for $\mathcal{D}_\texttt{GE}$ from \citet{Suresh2021} and $\mathcal{D}_\texttt{JQ1-3}$ from \citet{Liu2019HCOMP}. Full set of results included in Appendix Table~\ref{table:acc_results_full}.} The \textit{best} score for each dataset bolded. }\label{table:acc_results}
\end{table*}
To address \textbf{Q1}, i.e., whether mixtures of data features and labels in Stage 1 lead to better ground truth population estimates, Table~\ref{tab:kl_model_summary} shows the model name, hyperparameter values, and mean KL divergence between the cluster centroid  $\hat{y}_i$ and each item's empirical distribution $y_i$ of the best cluster model for each dataset. The best choice for $w$ varies considerably across the datasets. The two datasets, $\mathcal{D}_{GE,JQ3}$ with the largest number of choices (28 and 12, respectively) both selected models with $w=0$, i.e., the label distributions alone provided the best results. This was somewhat surprising, especially considering that in both cases the number of annotators per item is less than the number of label choices. We suspected that such sparse distributions would be too noisy to learn from. But apparently the size of these label spaces alone leads to a rich, meaningful signal. 

On the other extreme, the dataset with the fewest annotators ($\mathcal{D}_{SI}$) per item selected a model with $w = 1$, i.e., it used only item features, and not the label distributions, to determine the clusters. This is what we would expect whenever there is relatively low confidence in the label distributions, which should be the case with so few labels per item.  Interestingly, it was the only dataset that did not select NBP (K-Means).

\begin{table*}[ht!]
\tiny
\centering
 \resizebox{\textwidth}{!}{
\begin{tabular}{llc|c|cccccccccccc}



  & \textbf{Post}                       & \textbf{Model}  & \textbf{KL}  & hired & fired & quitting & other way & raise & hours & complains & support & going & home  & none  & other  \\ \hline
$\mathcal{D}_\texttt{JQ3}$E1 & Thank you Alice for all             & Annotations              &              & 0     & 0     & 0        & 0         & 0     & 0     & 5         & 1       & 0     & 0     & 4     & 0      \\
      & the attention u caused              & CO-FMM-CNN-$0$     & 0.706 &  0.044 & 0.003 & 0.009 & 0.009 & 0.009 & 0.015 & 0.208 & 0.017 & 0.060 & 0.042 & 0.318 & 0.265    \\ 
      & today at work                       & CO-FMM-CNN-$1$   & 1.11         & 0.07  & 0.063 & 0.136    & 0.084     & 0.091 & 0.002 & 0.293     & 0.019   & 0.019 & 0.043 & 0.071 & 0.098  \\
      &                                     & CO-NBP-CNN-$0.75$   & 0.63         & 0.05  & 0.082 & 0.062    & 0.023     & 0.048 & 0.005 & 0.382     & 0.056   & 0.011 & 0.021 & 0.134 & 0.123  \\ \hline
$\mathcal{D}_\texttt{JQ3}$E2 & Going to work 4PM to &  &      & \multicolumn{12}{c}{}\\ 
 & 12AM is NOT what I                 & Annotations &      & 0 &  0 &  1 &  0 &  1 &  1 &  4 &  1 &  5 &  0 &  1 &  0\\ 
      & want to do.. I have my & CO-FMM-CNN-$0$    & 0.597        & 0.028 & 0.000 & 0.019    & 0.009     & 0.019 & 0.038 & 0.323     & 0.028   & 0.118 & 0.192 & 0.157 & 0.064  \\
      & black sweatpants & CO-FMM-CNN-$1$ & 1.860        & 0.028 & 0.047 & 0.148    & 0.000     & 0.000 & 0.000 & 0.220     & 0.000   & 0.380 & 0.050 & 0.127 & 0.000  \\
&  spread out, though &  CO-NBP-CNN-$0.75$   & 0.522        & 0.002 & 0.047 & 0.138    & 0.000     & 0.039 & 0.021 & 0.220     & 0.001   & 0.244 & 0.080 & 0.207 & 0.001  \\
\hline
\end{tabular}

}

  

\caption{\small{Two examples from $\mathcal{D}_{JQ3}$. In the first example the author's sarcasm is missed by 4 out of 10 annotators who label the comment as \emph{none of the above but job related} and in the second, a similar sentiment is labeled as \emph{going to work} when \emph{hours} or \emph{complaining about work} are chosen by others. The act of ``laying out [work] clothes'' was not noted by many annotators.}}
\vspace{-0.5cm}
\label{tab:jbexample}

\end{table*}
In general, the mean KL-divergence for all selected models was quite low, suggesting that the items clustered together tended to have very similar label distributions. One might expect for there to be more divergence the higher $w$ is, because clustering with higher $w$ relies less directly on the label distributions.  But, reading across the the results, there does not appear to be any relationship between $w$ and KL-divergence. The datasets themselves are very different from one another, and so perhaps it is unlikely that something as simple as the mixing parameter $w$ would change the final label assignment.

For \textbf{Q2}, i.e., whether mixtures of data features and labels in Stage 1 improve the label distribution prediction in Stage 2, we measure the mean $\mbox{KL}(y_i \| \mathcal{H}(x_i))$,  where $\mathcal{H}$ is one of the supervised learning models trained on each of the clustering models.  For all datasets, the best cluster-based models in Table \ref{table:KL_results} outperform the baselines from Table~\ref{table:KL_results}. Among the clustering models, as with \textbf{Q1} there is a lot of variation among which values for $w$ give the best performance. But while the differences appear significant, they are not substantial, suggesting that subtle differences in the data or the inductive biases of particular clustering models are driving the variance. 

It is interesting to note that \textbf{DS+CNN} is always close to the worst model and often the worst by far. This may be because (a) that model treats disagreement as a sign of poor annotation and seeks to eliminate it, whereas our model is designed to preserve disagreement (b) \textbf{DS} models individual annotator-item pairs and the datasets we study here (which are representative of most datasets currently available) have very sparse label sets, and so overfitting is a concern.

For \textbf{Q3}, Table \ref{table:KL_results} (bottom) shows the classification prediction results, where evaluation is measured by accuracy, i.e., the proportion of test cases where the $\arg\max$ label of the (ground truth) training input label distribution is equal to that of the $\arg\max$ predicted label distribution. Here the results are mixed between the non-clustering (Table \ref{table:acc_results}) and clustering (Table \ref{table:acc_results}) models, and the variation in terms of significance and substance is in line with \textbf{Q1}. Once again, \textbf{DS+CNN} is the overall worst performer, even though here the goal is single-label inference, i.e., exactly what \textbf{DS} is designed for. 

\section{Discussions and Ethical Considerations}

Our results for \textbf{Qs 2--3} show that cluster-based aggregation universally improves the performance of distributional learning. This seems to confirm that clustering is a powerful tool for combating label sparseness to predict population-level annotator responses. However, results were mixed for single-label learning.  Also, among the clustering methods in both distributional and single-label learning, there was relatively little variance in performance as $w$ varies. 
\begin{figure*}[t]
\begin{subfigure}[t]{0.48\textwidth}
\begin{center}
\includegraphics[width=0.45\textwidth]{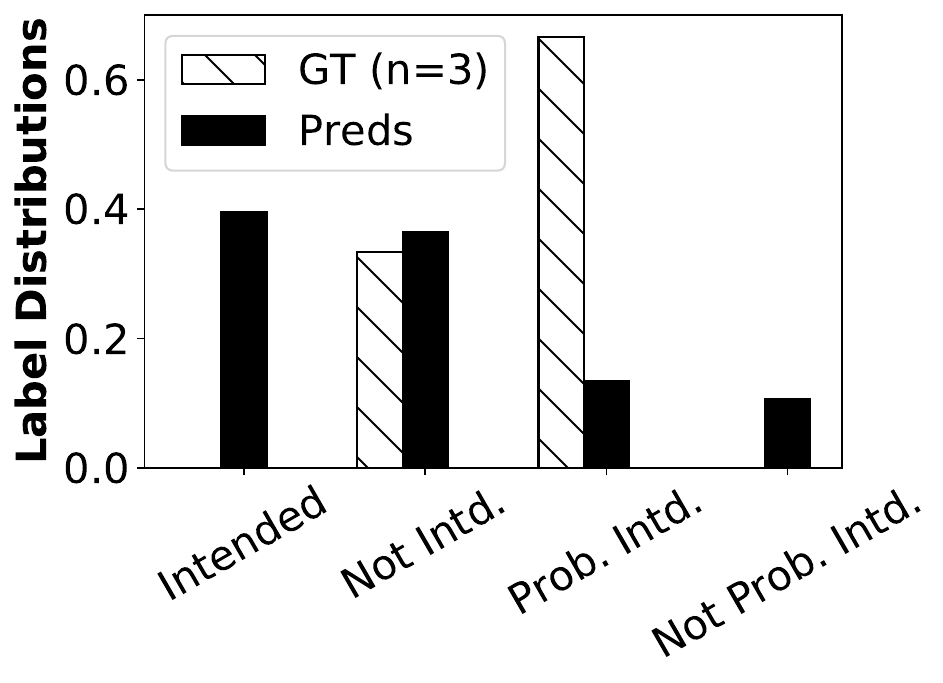}
\small
\begin{tabular}{c|ccccc}
\multicolumn{5}{c}{Annotator Demographics} \\
\textbf{Label} &\textbf{Gender} &\textbf{PL} &\textbf{Race} &\textbf{Age} \\ \hline
PI &Woman & Mod. Cons. &Hispanic &33 \\
PI &Woman & Mod. Cons. &White &35 \\
NI &Woman & Liberal &White &18 \\
\end{tabular}
    \caption{$\mathcal{D}_\texttt{SI}$E4 ``BPD is a genetic condition caused by having two x chromosomes. When a man is diagnosed with BPD it's just a professional way of saying he's acting like ae c***''. Source: r/Incels}
    \label{fig:SI_example4}
\end{center}
\end{subfigure}\hfill
\begin{subfigure}[t]{0.48\textwidth}
\begin{center}
\includegraphics[width=0.45\textwidth]{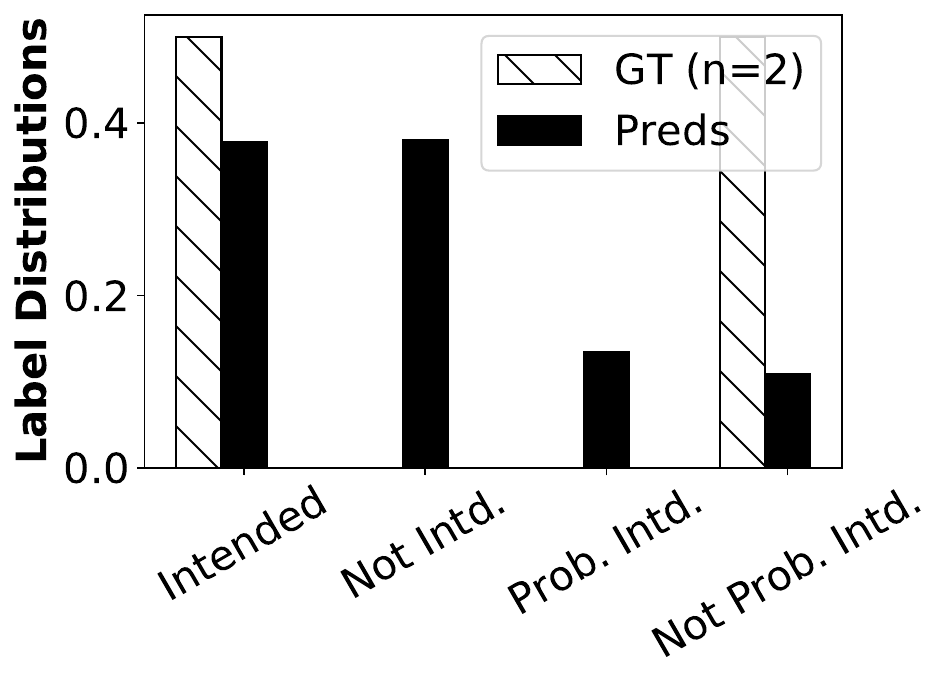}
\small
\begin{tabular}{c|ccccc}
\multicolumn{5}{c}{Annotator Demographics} \\
\textbf{Label} &\textbf{Gender}  &\textbf{PL} &\textbf{Race} &\textbf{Age} \\ \hline
NPI &Man &Liberal &White &29 \\
I &Man  &Liberal &White &39 \\
\end{tabular}
    \caption{$\mathcal{D}_\texttt{SI}$E5 ``Why should you never rape elderly women? because sex is for procreation you f****** Heathen!'' Source: r/darkjokes}
    \label{fig:SI_example5}
\end{center}
 \end{subfigure}
 
 \begin{subfigure}[t]{0.48\textwidth}
\begin{center}

\includegraphics[width=0.45\textwidth]{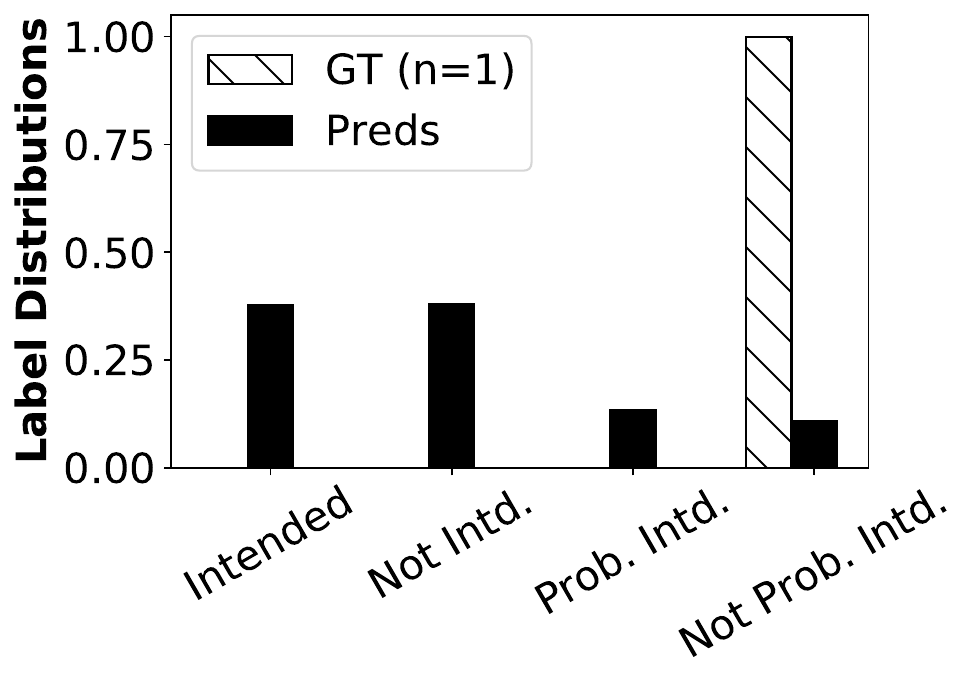}
\small
\begin{tabular}{c|ccccc}
\multicolumn{5}{c}{Annotator Demographics} \\
\textbf{Label} &\textbf{Gender} &\textbf{PL} &\textbf{Race} &\textbf{Age} \\ \hline
NPI &Man &Other &White &28 \\
\end{tabular}
    \caption{$\mathcal{D}_\texttt{SI}$E6 ``the klu klux klan. [repeat] the original boys in the hood''.  Source: r/meanjokes}
    \label{fig:SI_example6}
\end{center}
\end{subfigure}\hfill
\begin{subfigure}[t]{0.48\textwidth}
\begin{center}
\includegraphics[width=0.45\textwidth]{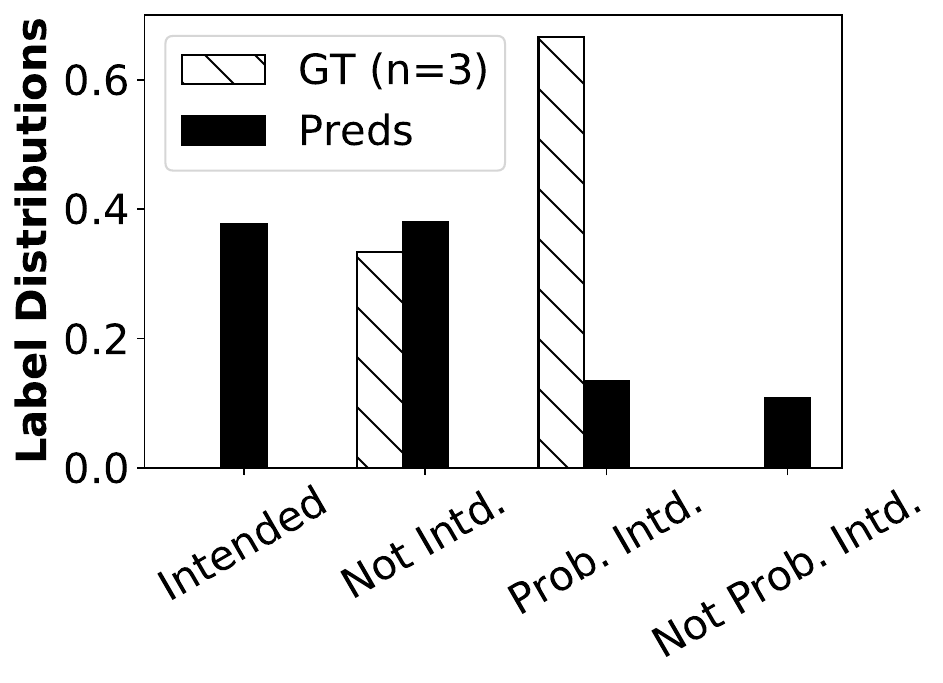}
\small
\begin{tabular}{c|ccccc}
\multicolumn{5}{c}{Annotator Demographics} \\
\textbf{Label} &\textbf{Gender} &\textbf{PL} &\textbf{Race} &\textbf{Age} \\ \hline
PI &Man &Moderate Liberal &White &35 \\
PI &Woman &Liberal &Asian &25 \\
NI &Man &Liberal &White &35 \\
\end{tabular}
    \caption{$\mathcal{D}_\texttt{SI}$E7 ``She can't regret it afterwards if she isn't breathing.'' Source: Gab}
    \label{fig:SI_example7}
\end{center}
 \end{subfigure}

    \caption{Examples from $\mathcal{D}_{SI}$ \cite{sap2019social}, human annotations (GT, striped bar) and predictions from the CO-FMM-CNN-$1$ model (Preds, solid bar). Here $n$ = number of human annotators, Mod. Cons. = moderate conservative, PL = political leaning, I = intended, NI = not intended, PI = probably intended, and  NPI = not probably intended.}
 \label{fig:SI_pred_examples}
 \vspace{-0.5cm}
 \end{figure*}
\begin{figure}[h]
    \centering
    \includegraphics[width=0.8\columnwidth]{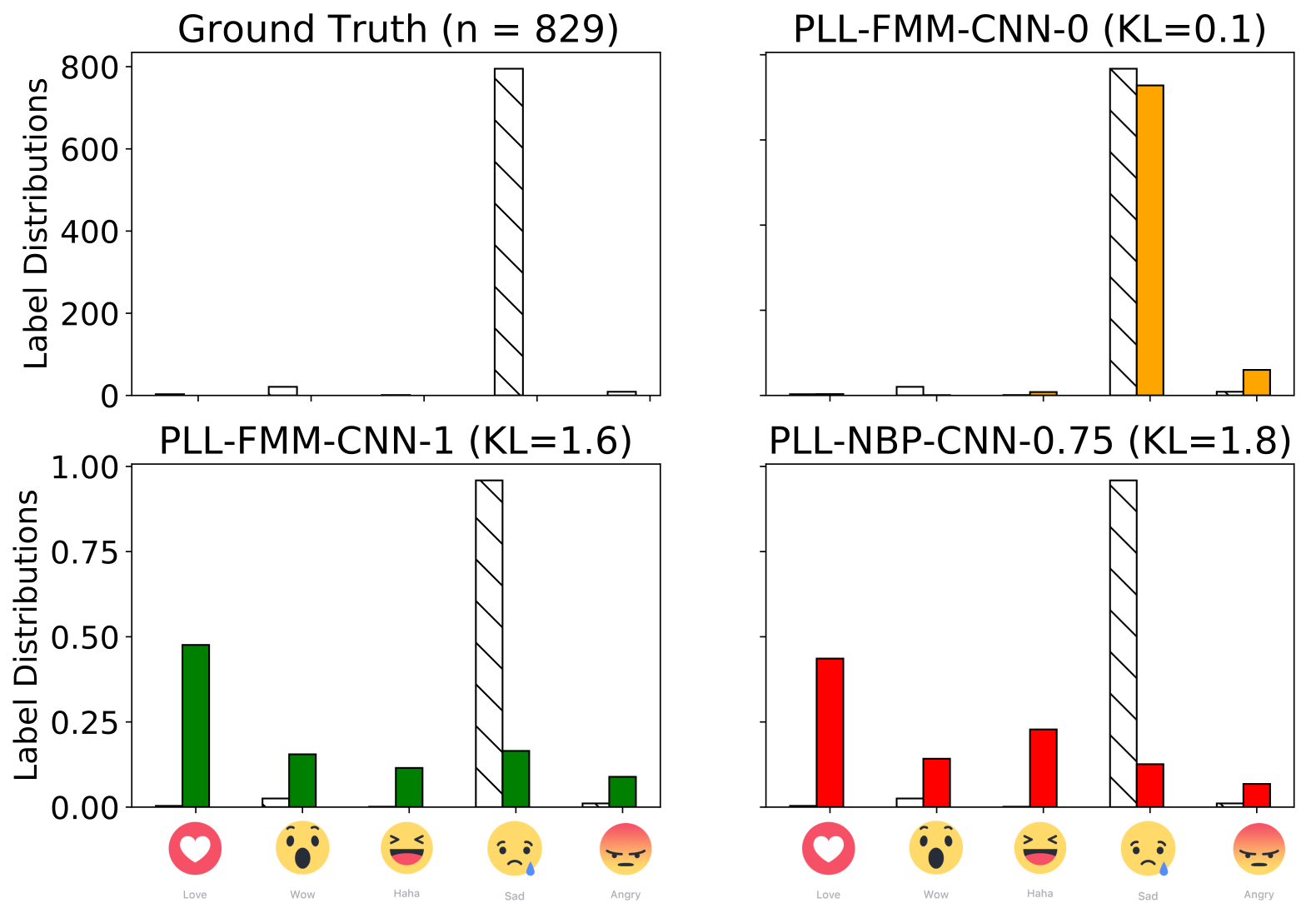}
    \caption{Example post, $\mathcal{D}_{FB}$E4 post: ``[i]t has been now about 15 hours since the child taken into water, so we know that we are working on recovering''. Here $n$ denotes number of human annotators and $KL$ is the KL-divergence when evaluated against the empirical ground truth. The striped bar denotes the human annotations. Reactions are \textit{love, wow, haha, sad,} and \textit{angry.}}
    \label{fig:fb_full_example}
    \vspace{-0.5cm}
\end{figure}

The latter is certainly a negative result with respect to the technical AI question of whether or not to use both data features and label distributions in cases when we \emph{do} cluster. But it is positive in that, combined with the overall superior performance of clustering for population-level learning, it shows that \emph{either} label features or label distributions are adequate for realizing the benefits of clustering as a means of label distribution regularization. It also suggests that annotator disagreements are, in fact, meaningful and essential.

To gain a better sense of how these methods can be used to address annotator inequality, we extract examples from $\mathcal{D}_\texttt{JQ3}$ (Table~\ref{tab:jbexample}), $\mathcal{D}_{FB}$ (Figure~\ref{fig:fb_full_example}), and $\mathcal{D}_{SI}$ (Figure~\ref{fig:SI_pred_examples}). We select examples from among the data items with the lowest KL-divergence scores between their empirical label distributions and their predictions according to the CO-FMM-CNN-$0$ model. We report their predicted distributions according to this model and two other models at a data item level.

Here, we see that the predicted distributions seem to differ from the empirical distributions and each other in meaningful ways. This is because our models rely on other items with similar label distributions or language to normalize reactions.  For instance, in example $\mathcal{D}_\texttt{FB}$E4, we see that the heavy annotator response to sad (795 responses) is retained when $w = 0$ (0.910), when only labels determine the clusters, but it decreases dramatically (to $0.165$ and $0.126$) as $w$ increases. These examples show that when we introduce text into the clustering phase, the overall performance may not change, but qualitative differences may be quite significant at the item level.


The examples in Figure~\ref{fig:SI_pred_examples} were surfaced by randomly sampling Reddit $\mathcal{D}_\texttt{SI}$ for posts whose predictions, using our models, differed from the human annotation. These examples all elicit ways of interpreting social media posts that contrast model predictions, human annotator choices, and our observations about offensiveness and toxicity. 
Example $\mathcal{D}_\texttt{SI}$E4, (Figure~\ref{fig:SI_example4}) is an offensive joke that mocks women and people with a mental health disorder called borderline personality disorder (``BPD''). In contrast, the human annotation was split between \textit{not intended to be offensive} and \textit{probably intended to be offensive}. No human chose \textit{intended to be offensive}, yet our algorithm predicted it might be, reflecting the deniability that comes from phrasing offensive speech as a ``joke.''
 
Example $\mathcal{D}_\texttt{SI}$E5,  (Figure~\ref{fig:SI_example6}) is a joke about rape and older women. It is offensive because it associates rape with sex as opposed to rape with violence and sex with procreation. This is a challenging case for a typical ML classifier---there is no majority, and the label polarities are also opposite. In this case, our prediction correctly identifies the majority label. This may be due to our models grouping similar data items of similar content, supporting items such as this when there is contrasting confidence in human annotators.
 
Example $\mathcal{D}_\texttt{SI}$E6 (Figure~\ref{fig:SI_example5}) is offensive because it makes light of the hate group KKK wearing hoods by identifying them with an NWA song and film about African American teenagers (``boyz n the hood''). The PLL prediction also indicates that this post may have been \textit{intended to be offensive}. But the human annotator thought it was \textit{probably not intended to be offensive}. This is another case where our prediction aligns with our judgment.

Example $\mathcal{D}_\texttt{SI}$E7,  (Figure~\ref{fig:SI_example7}) is offensive because it alludes to a woman being dead and thus not having agency; it seems threatening. Two human annotators chose this to be \textit{probably intended to be offensive}, and one annotator considered it \textit{not intended to be offensive}. The prediction finds this \textit{intended to be offensive}.

A commonality among these examples is that they all contain an element of deniability---the poster can always claim they were only joking. One challenge with content moderation is where to draw the line. When does the potential harm of letting an offensive post through outweigh the winnowing of free discourse? The answer often depends on context. The population-level learning approach we advocate here can help provide a more nuanced view into annotator response. It may also provide context on opinions to inform decisions about what should and should not be censored.

Our work also supports the findings from \cite{DBLP:journals/corr/abs-2111-07997}, where they studied the underlying reasons why annotators disagree on subjective content, such as offensive language annotation. The examples show how the proposed models can identify offensive content even with unreliable training data (human annotations).

\section{Conclusion}
Human annotation is often an expensive-to-acquire, challenging, and subjective resource for supervised machine learning. The obstacles to using human decisions in ML classification tasks are even more apparent when the problem domain is social media content. The nuance, disagreement, and diversity of opinions by humans augment and enrich the complex decisions machine learning attempts to surface. To gain as much utility as possible from this valuable resource, we propose and subsequently \emph{CrowdOpinion} to retain these human judgments in the data prediction pipeline for as long as possible. First, this work introduces a novel method for mixing language features and label features into label distribution estimators to improve population-level learning. Then, we evaluated our approach against different baselines and experimented with datasets containing varying amounts of annotator disagreements. Our results suggest that (i) clustering is an effective measure for countering the problem of label sparseness when learning a population-level distribution of annotator responses, (ii) data features or label distributions are equally helpful as spaces in which to perform such clustering, and thus (iii) label distributions are meaningful signals that reflect the content of their associated items.

\section*{Limitations}

\noindent\textbf{Evaluation:} We evaluate work as a single-label learning problem (accuracy) and a probability distribution (KL). These metrics do not fully capture the nuances of the crowd \cite{inel_crowdtruth_2014}. We hope to build on this work by moving beyond general population-level predictions to predictions on subpopulations of interest, such as vulnerable communities. We hope to develop better methods for evaluating and assessing the performance of population-level learning. 

The range of mixing ($w=$) of the language features and labels in our experiments could be further delved into. Our experiments cover weights ranging from 0 to 100 in quartiles, but this parameter, as a hyperparameter, could benefit from additional experiments in finer ranges. 

\noindent\textbf{Datasets:} Our experimental datasets have been primarily in English. In addressing the ability to generalize, we hope to explore other offensive or hate speech-related datasets from other languages. The challenge of evaluating our models with other languages is acquiring a dataset with annotator-level labels, a rare resource for English datasets and challenging for other languages. Finally, we hope our methods open the discussion to building nuanced systems that capture human disagreement while studying subjective content on social media. 

\noindent\textbf{Computation:} As our experiments follow a two-stage setup, the first phase (data mixing) of it can be further optimized to run on GPUs similar to the second phase (classification), which is running on GPU through the TensorFlow/Keras implementation. The first phase utilizes libraries through Sckit-learn, BNPY, and scripts through Python (NBP), which can be a bottleneck for implementing the work and expanding.

\section*{Ethical Considerations}
Our analysis constitutes a secondary study of publicly available datasets and thus is considered exempt from a federal human subjects research perspective. However, as with any study that involves data collected from humans, there is a risk that it can be used to identify people \cite{hovy-spruit-2016-social,kralj_novak_handling_2022}. We understand these risks and train and test our models on anonymized data to minimize them. In addition, it is essential to note that any methods identifying marginalized voices can also aid in selective censorship. Our models in Stage 1 and Stage 2, generate rich soft label distributions, this can be helpful for ML models to learn from a representative label. The distributions can also help with making decisions taking into account the right to freedom of expression and right to safety for human content creators, consumers, and annotators. 

\section*{Acknowledgments}
The funding for this research was provided by a Google Research Award, along with support from Google Cloud Research credits. Additionally, resources from Research Computing at the \citet{ritrc} were utilized. We express our gratitude to the anonymous reviewers for their valuable feedback and suggestions on our work, as well as to the wider community for their support.

\bibliography{pldl,cyril_zotero}
\bibliographystyle{acl_natbib}

\appendix
 
\section{Dataset Sources}

\begin{enumerate}
    \item $\mathcal{D}_{GE}$ by \citet{demszky2020goemotions} - Available at \url{https://github.com/google-research/google-research/tree/master/goemotions}
    \item $\mathcal{D}_{JQ1-3}$ by \citet{P16-1099} - Available at  \url{https://github.com/Homan-Lab/pldl_data}
    \item $\mathcal{D}_{SI}$ by \citet{sap2019social} - Available at \url{https://homes.cs.washington.edu/~msap/social-bias-frames/index.html}
    \item $\mathcal{D}_{FB}$ available at \citet{WolfFb}
\end{enumerate}

\subsection{GoEmotions ($\mathcal{D}_\texttt{GE}$)} 
This is one of the largest, hate-speech related datasets of around 58,000 Reddit comments collected by \citet{demszky2020goemotions}. The comments are annotated by a total of 82 MTurkers with 27 emotions or ``neutral,'' yielding 28 annotation labels total: \emph{admiration, amusement, anger, annoyance, approval, caring, confusion, curiosity, desire, disappointment, disapproval, disgust, embarrassment, excitement, fear, gratitude, grief, joy, love, nervousness, optimism, pride, realization, relief, remorse, sadness, surprise,} and \emph{neutral}. The number of annotations per item varies from 1  to 16. 


\subsection{Jobs ($\mathcal{D}_\texttt{JQ1-3}$)}
\citet{P16-1099} asked five annotators each from MTurk and F8 platforms to label work related tweets according to three questions: point of view of the tweet ($\mathcal{D}_\texttt{JQ1}$: \emph{1st person}, \emph{2nd person}, \emph{3rd person}, \emph{unclear}, or \emph{not job related}), subject's employment status ($\mathcal{D}_\texttt{JQ2}$: \emph{employed}, \emph{not in labor force}, \emph{not employed}, \emph{unclear}, and \emph{not job-related}), and employment transition event ($\mathcal{D}_\texttt{JQ3}$: \emph{getting hired/job seeking}, \emph{getting fired}, \emph{quitting a job}, \emph{losing job some other way}, \emph{getting promoted/raised}, \emph{getting cut in hours}, \emph{ complaining about work}, \emph{offering support}, \emph{going to work}, \emph{coming home from work}, \emph{none of the above but job related}, and \emph{not job-related}).

\subsection{SBIC Intent ($\mathcal{D}_\texttt{SI}$)} 
The Social Bias Inference Corpus ($\mathcal{D}_\texttt{SI}$) dataset is made up of $\sim$45,000 posts from Reddit, Twitter, and hate sites collected by \citet{sap2019social}. It was annotated with respect to seven questions: offensiveness, intent to offend, lewdness, group implications, targeted group, implied statement, in-group language. Out of these predicates, we consider only the intent to offend question (as it had the richest label distribution patterns) with the label options: \emph{Intended}, \emph{Probably Intended}, \emph{Probably Not Intended}, and \emph{Not Intended}. The number of annotations per data item varies between 1 and 20 annotations. 


\subsection{Facebook ($\mathcal{D}_\texttt{FB}$)} 
The original multi-lingual dataset is Facebook posts written on the 144 most-liked pages during 4 months in 2016. The posts all come from pages hosted by news entities or public figures with a large fanbase interacting through comments and reactions. Each item consists of the post text (we remove all non-text data) and we take as the label set the (normalized) distribution of the post's reactions: \emph{like}, \emph{love}, \emph{haha}, \emph{wow}, \emph{sad}, and \emph{angry}. However, as \emph{like} tends to dominate, following \citet{tian2017facebook} we eliminate that reaction before we normalize. We perform language detection \footnote{Google Translate Language Detection \url{https://bit.ly/33g7Ct3}} and subsample 2,000 English-only posts. The annotations per item varies widely from 50 to 71,399. In contrast to other datasets, $\mathcal{D}_\texttt{FB}$ is a special case since annotations for it come from users of the social network. The users are ``reacting'' to a post in contrast to a human annotator annotating a post for a specified task. The randomness of users reacting to a post and posts being from different domains make it a special case. 

\section{Experimental Setup}
Our experimental setup consists of the following configurations; Setup \#1 - Ubuntu 18.04, Intel i6-7600k (4 cores) at 4.20GHz, 32GB RAM, and nVidia GeForce RTX 2070 Super 8GB VRAM. Setup \#2 - Debian 9.8, Intel Xeon (6 cores) at 2.2GHz, 32GB RAM, and nVidia Tesla P100 12GB VRAM. For a single pass through on a dataset, the estimated time of completion is 8 hours per language representation model on Setup \#2, which is the slowest out of the two.

In our experimental setup, we compare our language based models to other PLDL models based on annotations and baselines from prior research. For comparison sake, we built our own experimental setup similar to the models used by \citet{Liu2019HCOMP,Weerasooriya2020}. 

Experiments tracked with ``Weights and Biases`` by \citet{wandb}.

\section{Complete set of results for CO}
See Table~\ref{table:KL_results_full} for KL-Divergence and Table~\ref{table:acc_results_full} and for accuracy results.
\begin{table*}[h!]
\smaller
\centering
\setlength{\tabcolsep}{3pt}
\begin{tabular}{c|c|cccc}
\textbf{Data-} & \textbf{Baseline} & \multicolumn{4}{c}{\textbf{CO-$\mathcal{C}$-CNN-$w$}} \\
\textbf{set} &\textbf{$w=0$} &\textbf{$w=0.25$} &\textbf{$w=0.50$} &\textbf{$w=0.75$} &\textbf{$w=1$} \\
\hline
\multicolumn{6}{c}{$\mathcal{C} = $\textbf{FMM Clustering}} \\
\hline
$\mathcal{D}_\texttt{FB}$ &0.707$\pm$0.003 &\textbf{ 0.686$\pm$0.004} &0.687$\pm$0.004 &0.689$\pm$0.003 &\textbf{ 0.686$\pm$0.003} \\
$\mathcal{D}_\texttt{GE}$  &2.011$\pm$ 0.002 &2.010$\pm$0.001 &2.008$\pm$0.002 &2.005$\pm$0.001 &\textbf{ 2.004$\pm$0.002} \\
$\mathcal{D}_\texttt{JQ1}$ &\textbf{ 0.458$\pm$0.001} &0.464$\pm$0.007 &0.468$\pm$0.011 &0.46$\pm$0.004 &0.461$\pm$0.006 \\
$\mathcal{D}_\texttt{JQ2}$ &\textbf{ 0.515$\pm$0.001} &0.522$\pm$0.009 &0.517$\pm$0.005 & 0.515$\pm$0.003 &0.518$\pm$0.007 \\
$\mathcal{D}_\texttt{JQ3}$  &\textbf{ 0.887$\pm$0.001} &0.892$\pm$0.004 &0.889$\pm$0.005 &0.889$\pm$0.003 &0.890$\pm$0.003 \\
$\mathcal{D}_\texttt{SI}$  &0.991$\pm$0.003 &0.992$\pm$0.005 &0.993$\pm$0.003 &0.927$\pm$0.027 &\textbf{ 0.86$\pm$0.026} \\
\hline
\multicolumn{6}{c}{$\mathcal{C} = $\textbf{GMM Clustering}} \\
\hline
$\mathcal{D}_\texttt{FB}$ &0.684$\pm$0.001 &0.683$\pm$0.003 &0.682$\pm$0.001 &\textbf{ 0.680$\pm$0.001} &0.685$\pm$0.002 \\
$\mathcal{D}_\texttt{GE}$ &1.999$\pm$ 0.001 &\textbf{ 1.998$\pm$0.001} &2.002$\pm$0.006 &2.000$\pm$0.003 &\textbf{ 1.998$\pm$ 0.003} \\
$\mathcal{D}_\texttt{JQ1}$ &0.450$\pm$0.001 &0.467$\pm$0.001 &0.447$\pm$0.004 &0.437$\pm$0.001 &\textbf{ 0.427$\pm$0.01} \\
$\mathcal{D}_\texttt{JQ2}$ &0.513$\pm$0.002 &0.512$\pm$0.001 &\textbf{ 0.510$\pm$0.003} &0.514$\pm$0.001 &0.516$\pm$0.004 \\
$\mathcal{D}_\texttt{JQ3}$ &0.880$\pm$0.001 &0.881$\pm$0.001 &\textbf{ 0.870$\pm$0.001} &0.885$\pm$0.001 &0.889$\pm$0.005 \\
$\mathcal{D}_\texttt{SI}$ &0.882$\pm$0.008 &\textbf{ 0.877$\pm$0.024} &0.904$\pm$0.021 &0.9$\pm$0.031 &0.894$\pm$0.026 \\
\hline
\multicolumn{6}{c}{$\mathcal{C} =$ \textbf{K-Means clustering}} \\
\hline
$\mathcal{D}_\texttt{FB}$ &\textbf{ 0.680$\pm$0.0} &0.687$\pm$0.001 &\textbf{ 0.680$\pm$0.001} &0.688$\pm$0.001 &0.684$\pm$0.0 \\
$\mathcal{D}_\texttt{GE}$ &\textbf{ 1.998$\pm$0.001} &1.999$\pm$0.002 &2.002$\pm$0.006 &2.001$\pm$0.004 &2.000$\pm$0.004 \\
$\mathcal{D}_\texttt{JQ1}$ &0.457$\pm$0.001 &0.456$\pm$0.0 &0.457$\pm$0.001 &0.447$\pm$0.001 &\textbf{ 0.434$\pm$0.001} \\
$\mathcal{D}_\texttt{JQ2}$ &\textbf{ 0.499$\pm$0.001} &0.510$\pm$0.001 &0.510$\pm$0.002 &0.512$\pm$0.002 &0.513$\pm$0.001 \\
$\mathcal{D}_\texttt{JQ3}$ &0.874$\pm$0.001 &0.883$\pm$0.001 &\textbf{ 0.853$\pm$0.001} &0.888$\pm$0.001 &0.889$\pm$0.001 \\
$\mathcal{D}_\texttt{SI}$ &\textbf{ 0.857$\pm$0.008} &0.886$\pm$0.024 &0.889$\pm$0.028 &0.895$\pm$0.028 &0.894$\pm$0.027 \\
\hline
\multicolumn{6}{c}{$\mathcal{C} = $ \textbf{LDA Clustering}} \\
\hline
$\mathcal{D}_\texttt{FB}$ & 0.684$\pm$0.0 &\textbf{ 0.683$\pm$0.0} &0.684$\pm$0.0 &0.684$\pm$0.0 &0.684$\pm$0.0 \\
$\mathcal{D}_\texttt{GE}$ &\textbf{ 1.987$\pm$0.0} &1.997$\pm$0.0 &1.995$\pm$0.0 &1.999$\pm$0.002 &1.999$\pm$0.001 \\
$\mathcal{D}_\texttt{JQ1}$ &0.458$\pm$0.001 &0.457$\pm$0.001 &\textbf{ 0.456$\pm$0.001} &0.459$\pm$0.001 &0.458$\pm$0.001 \\
$\mathcal{D}_\texttt{JQ2}$ &\textbf{ 0.512$\pm$0.0} &0.514$\pm$0.001 &0.515$\pm$0.0 &0.513$\pm$0.001 &\textbf{ 0.512$\pm$0.001} \\
$\mathcal{D}_\texttt{JQ3}$ &0.884$\pm$0.0 &0.885$\pm$0.0 &0.880$\pm$0.001 &0.834$\pm$0.0 &\textbf{ 0.823$\pm$0.0} \\
$\mathcal{D}_\texttt{SI}$ &0.932$\pm$0.0 &0.980$\pm$0.0 &0.92$\pm$0.045 &\textbf{ 0.867$\pm$0.018} &0.905$\pm$0.023 \\
\hline
\multicolumn{6}{c}{$\mathcal{C} = $\textbf{NBP Pooling}} \\
\hline
$\mathcal{D}_\texttt{FB}$ &0.688$\pm$0.003 &\textbf{ 0.686$\pm$0.001} &0.687$\pm$0.002 &0.688$\pm$0.004 &0.69$\pm$0.007 \\
$\mathcal{D}_\texttt{GE}$  &2.002$\pm$0.005 &\textbf{ 2.0$\pm$0.002} &2.001$\pm$0.005 &2.001$\pm$0.001 &2.010$\pm$0.003 \\
$\mathcal{D}_\texttt{JQ1}$  &0.469$\pm$0.009 &0.485$\pm$0.026 &0.479$\pm$0.021 &0.475$\pm$0.012 &\textbf{ 0.457$\pm$0.0} \\
$\mathcal{D}_\texttt{JQ2}$ &0.520$\pm$0.007 &0.519$\pm$0.01 &0.519$\pm$0.007 &0.522$\pm$0.01 &\textbf{ 0.513$\pm$0.001} \\
$\mathcal{D}_\texttt{JQ3}$ &0.897$\pm$0.012 &0.889$\pm$0.005 &0.894$\pm$0.006 &0.889$\pm$0.007 &\textbf{ 0.883$\pm$0.0} \\
$\mathcal{D}_\texttt{SI}$ &0.900$\pm$0.024 &0.895$\pm$0.025 &0.894$\pm$0.028 &0.890$\pm$0.019 &\textbf{ 0.889$\pm$0.027} \\
\hline
\end{tabular}
 \caption{\small{KL-divergence results for the CO-$\mathcal{C}$-CNN-$w$ models from Algorithm \ref{alg:clpldl}, using various choices for clustering $\mathcal{C}$ and feature-label mixing $w$. Here $w=0$ is the baseline from \citet{Liu2019HCOMP} that uses label distributions in the clustering stage, and $w=1$ means that only data feature are used.  The \textit{best} score is bolded. Baseline from \citet{Liu2019HCOMP}.}}\label{table:KL_results_full}
\vspace{-0.3cm}
\end{table*}

\begin{table*}[h!]
\centering
\smaller
\setlength{\tabcolsep}{1pt}
\begin{tabular}{c|c|cccc}
\textbf{Data-} &\textbf{Baseline} &\multicolumn{4}{c}{\textbf{CO-$\mathcal{C}$-CNN-$w$}} \\
\textbf{set}  &$w=0$ &$w=0.25$ &$w=0.50$ &$w=0.75$ &$w=1$ \\
\hline
\multicolumn{6}{c}{\textbf{$\mathcal{C} = $ FMM Clustering}} \\
\hline
$\mathcal{D}_\texttt{FB}$  &0.780$\pm$0.001 &0.777$\pm$0.010 & 0.789$\pm$0.001 &0.787$\pm$0.001 &\textbf{ 0.790$\pm$0.001} \\
$\mathcal{D}_\texttt{GE}$ &0.949$\pm2e^{-16} $ &0.949$\pm2e^{-16}$ &0.923$\pm2e^{-16} $ &0.910$\pm2e^{-16} $ &0.948$\pm2e^{-16} $ \\
$\mathcal{D}_\texttt{JQ1}$ &\textbf{ 0.892$\pm$0.0} &0.890$\pm$0.0 &0.878$\pm$0.0 &0.880$\pm$0.0 &\textbf{ 0.892$\pm$0.0} \\
$\mathcal{D}_\texttt{JQ2}$ &\textbf{ 0.890$\pm$0.0} &0.812$\pm$0.0 &\textbf{ 0.890$\pm$0.0} &0.870$\pm$0.0 &0.830$\pm$0.0 \\
$\mathcal{D}_\texttt{JQ3}$ &0.878$\pm$0.002 &0.880$\pm$0.002 &0.870$\pm$0.003 & 0.881$\pm$0.002 &0.880$\pm$0.002 \\
$\mathcal{D}_\texttt{SI}$ &0.949$\pm$0.0 &\textbf{ 0.950$\pm$0.0} &0.940$\pm$0.0 &0.941$\pm$0.0 &0.942$\pm$0.0 \\
\hline
\multicolumn{6}{c}{\textbf{$\mathcal{C} = $ GMM Clustering}} \\
\hline
$\mathcal{D}_\texttt{FB}$ &0.785$\pm$0.001 &0.789$\pm$0.001 &0.787$\pm$0.001 &\textbf{ 0.798$\pm$0.001} &0.783$\pm$0.001 \\
$\mathcal{D}_\texttt{GE}$ &0.940$\pm$0.001 &0.949$\pm$ 0.001 &0.942$\pm$0.006 &0.949$\pm$0.003 &0.950$\pm$0.003 \\
$\mathcal{D}_\texttt{JQ1}$ &0.891$\pm1e^{-16}$ &0.888$\pm1e^{-16}$ &0.880$\pm1e^{-16}$ &\textbf{ 0.901$\pm1e^{-16}$} &0.890$\pm$0.0 \\
$\mathcal{D}_\texttt{JQ2}$ &0.870$\pm1e^{-16}$ &\textbf{ 0.875$\pm1e^{-16}$} &0.865$\pm1e^{-16}$ &0.800$\pm1e^{-16}$ &0.801$\pm$0.0 \\
$\mathcal{D}_\texttt{JQ3}$ &0.880$\pm$0.002 &0.881$\pm1e^{-16}$ &0.875$\pm$0.001 &0.870$\pm$0.002 &0.871$\pm$0.002 \\
$\mathcal{D}_\texttt{SI}$ &\textbf{ 0.949$\pm$0.0} &0.947$\pm$0.0 &0.945$\pm$0.0 &0.944$\pm$0.0 &0.943$\pm$0.0 \\
\hline
\multicolumn{6}{c}{\textbf{$\mathcal{C} = $ K-Means Clustering}} \\
\hline
$\mathcal{D}_\texttt{FB}$ &0.780$\pm$0.001 &0.783$\pm$0.001 & 0.786$\pm$0.001 &0.773$\pm$0.001 &0.765$\pm$0.001 \\
$\mathcal{D}_\texttt{GE}$ &0.940$\pm$0.000 &0.930$\pm$0.000 &0.930$\pm$0.000 &0.902$\pm$0.000 &0.938$\pm$0.000 \\
$\mathcal{D}_\texttt{JQ1}$ &0.890$\pm$0.0 &0.891$\pm$0.0 &\textbf{ 0.893$\pm$0.0} &0.890$\pm$0.0 &0.870$\pm$0.0 \\
$\mathcal{D}_\texttt{JQ2}$ &0.873$\pm$0.0 &0.870$\pm$0.0 &\textbf{ 0.875$\pm$0.0} &0.872$\pm$0.0 &0.870$\pm$0.0 \\
$\mathcal{D}_\texttt{JQ3}$  &0.881$\pm$0.0 &0.878$\pm$0.0 &0.875$\pm$0.0 &0.870$\pm$0.0 &0.830$\pm$0.001 \\
$\mathcal{D}_\texttt{SI}$ &0.775 $\pm$0.008 &\textbf{ 0.777$\pm$0.007} &0.76$\pm$0.028 &0.773$\pm$0.009 &0.759$\pm$0.023 \\
\hline
\multicolumn{6}{c}{\textbf{$\mathcal{C} = $ LDA Clustering}} \\
\hline
$\mathcal{D}_\texttt{FB}$ &0.784$\pm$0.0 &0.782$\pm$0.0 &0.787$\pm$0.0 &0.788$\pm$0.0 &0.789$\pm$0.0 \\
$\mathcal{D}_\texttt{GE}$ &0.949$\pm$0.0 &0.930$\pm$0.0 &0.935$\pm$0.0 &0.932$\pm$0.0 &0.950$\pm$0.0 \\
$\mathcal{D}_\texttt{JQ1}$ &0.891$\pm$0.0 &\textbf{0.893$\pm$0.0} &0.890$\pm$0.0 &0.891$\pm$0.0 &0.891$\pm$0.0 \\
$\mathcal{D}_\texttt{JQ2}$ &0.873$\pm$0.0 &0.875$\pm$0.0 &0.870$\pm$0.0 &0.878$\pm$0.0 &\textbf{ 0.879$\pm$0.0} \\
$\mathcal{D}_\texttt{JQ3}$ &0.880$\pm$0.0 &0.881$\pm$0.0 &0.882$\pm$0.0 &0.883$\pm$0.0 &0.879$\pm$0.001 \\
$\mathcal{D}_\texttt{SI}$ &0.932$\pm$0.0 &\textbf{ 0.980$\pm$0.0} &0.92$\pm$0.045 &0.867$\pm$0.018 &0.905$\pm$0.023 \\
\hline
\multicolumn{6}{c}{\textbf{$\mathcal{C} = $ NBP Clustering}} \\
\hline
$\mathcal{D}_\texttt{FB}$ &0.785$\pm$0.0 &0.781$\pm$0.0 &0.780$\pm$0.0 & 0.787$\pm$ 0.0 &0.785$\pm$0.0 \\
$\mathcal{D}_\texttt{GE}$ &0.850$\pm$0.0 &0.820$\pm$0.0 &0.810$\pm$0.0 &0.800$\pm$0.0 &0.805$\pm$0.0 \\
$\mathcal{D}_\texttt{JQ1}$ &0.890$\pm$0.0 &0.879$\pm$0.0 &0.890$\pm$0.0 &0.789$\pm$0.005 &\textbf{ 0.892$\pm$0.0} \\
$\mathcal{D}_\texttt{JQ2}$ &0.873$\pm$0.0 &\textbf{ 0.897$\pm$0.0} &0.880$\pm$0.0 &0.820$\pm$0.0 &0.865$\pm$0.0 \\
$\mathcal{D}_\texttt{JQ3}$ &0.880$\pm$0.002 &0.879$\pm$0.002 &0.865$\pm$0.002 &0.879$\pm$0.002 &0.881$\pm$0.0 \\
$\mathcal{D}_\texttt{SI}$ &0.755$\pm$0.036 &\textbf{ 0.767$\pm$0.019} &0.758$\pm$0.034 &0.761$\pm$0.016 &0.762$\pm$0.025 \\
\hline
\end{tabular}
 \caption{\small{Accuracy results for the CO-$\mathcal{C}$-CNN-$w$ models from Algorithm \ref{alg:clpldl}, using various choices for clustering $\mathcal{C}$ and feature-label mixing $w$. Here $w=0$ is the baseline from \citet{Liu2019HCOMP} that uses label distributions in the clustering stage, and $w=1$ means that only data feature are used. Since accuracy is a non-distributional statistic, we use the most frequent label for inference (though not during training; we use the same trained models as in Table \ref{tab:kl_model_summary}). The \textit{best} score is bolded. Baseline from \citet{Liu2019HCOMP}.}}\label{table:acc_results_full}

\end{table*}

\section{Entropy distributions}
\begin{figure*}[t!]
\begin{minipage}{0.33\linewidth}
\includegraphics[width=\linewidth]{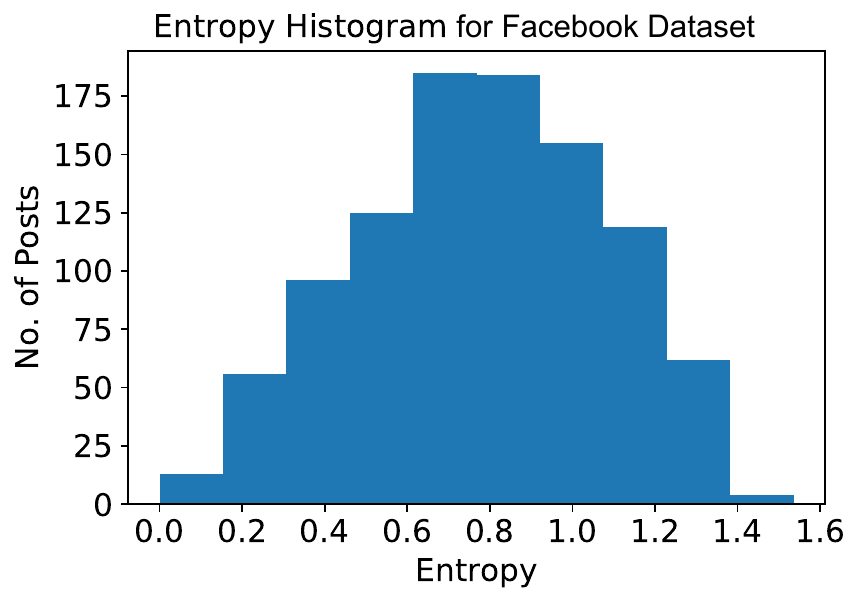}
\end{minipage} 
\begin{minipage}{0.33\linewidth}
    \includegraphics[width=\linewidth]{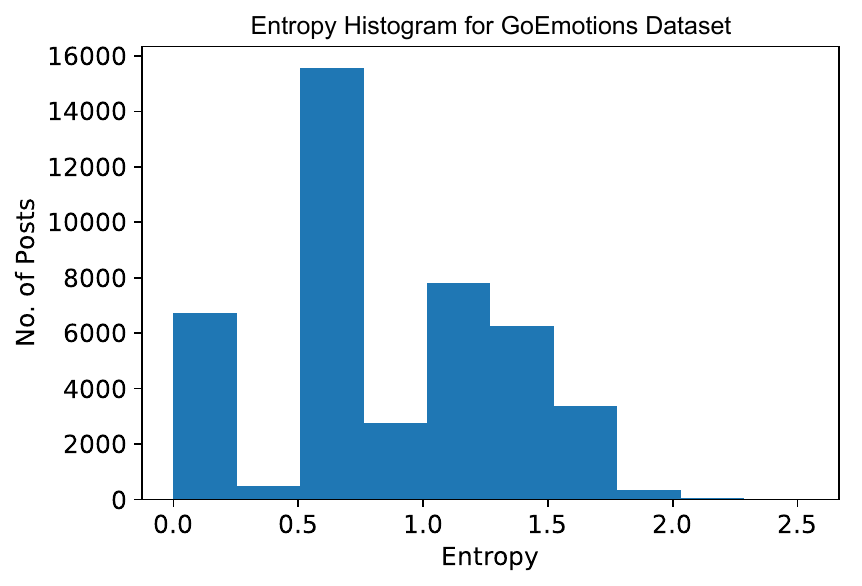}
\end{minipage}%
\begin{minipage}{0.33\linewidth}
    \includegraphics[width=\linewidth]{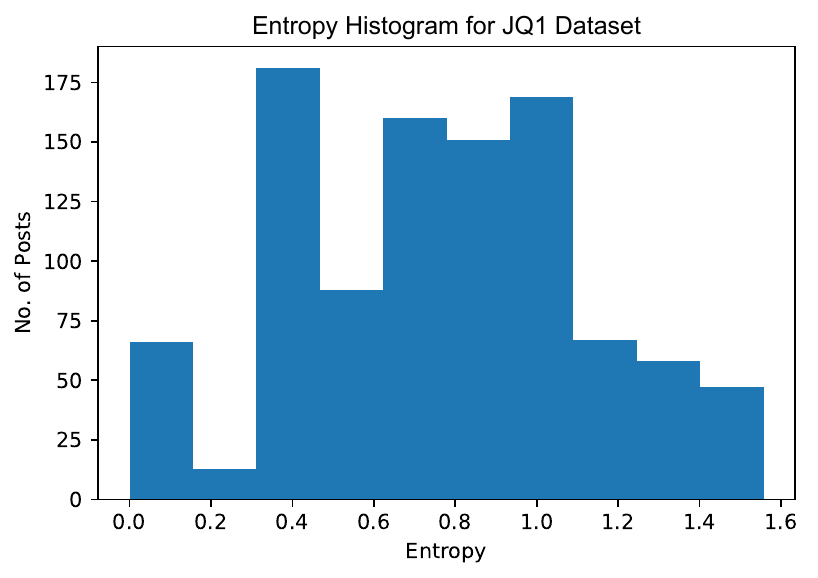}
\end{minipage}%
\hfill
\begin{minipage}{0.33\linewidth}
\includegraphics[width=\linewidth]{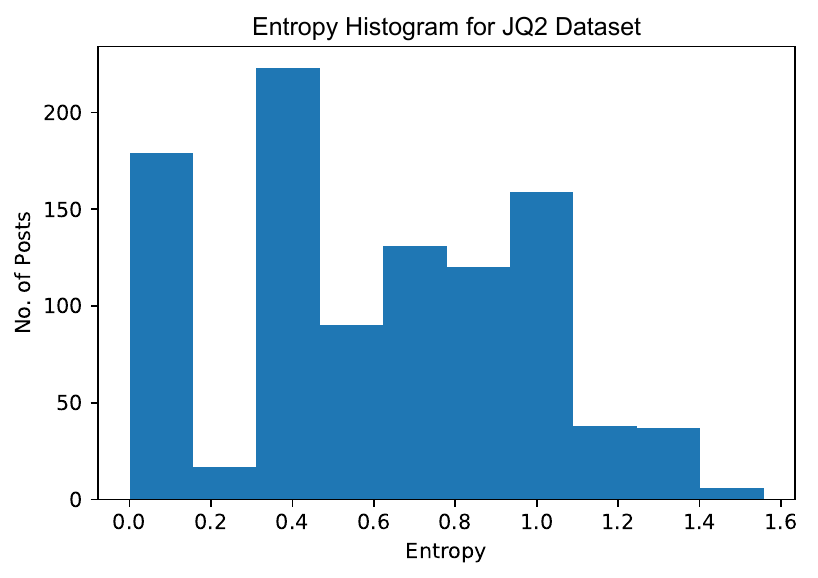}
\end{minipage} 
\begin{minipage}{0.33\linewidth}
    \includegraphics[width=\linewidth]{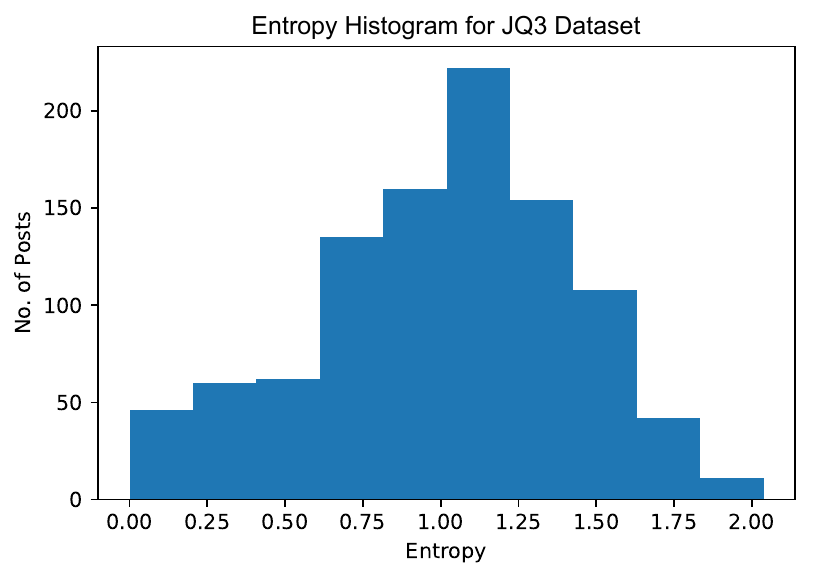}
\end{minipage}%
\begin{minipage}{0.33\linewidth}
    \includegraphics[width=\linewidth]{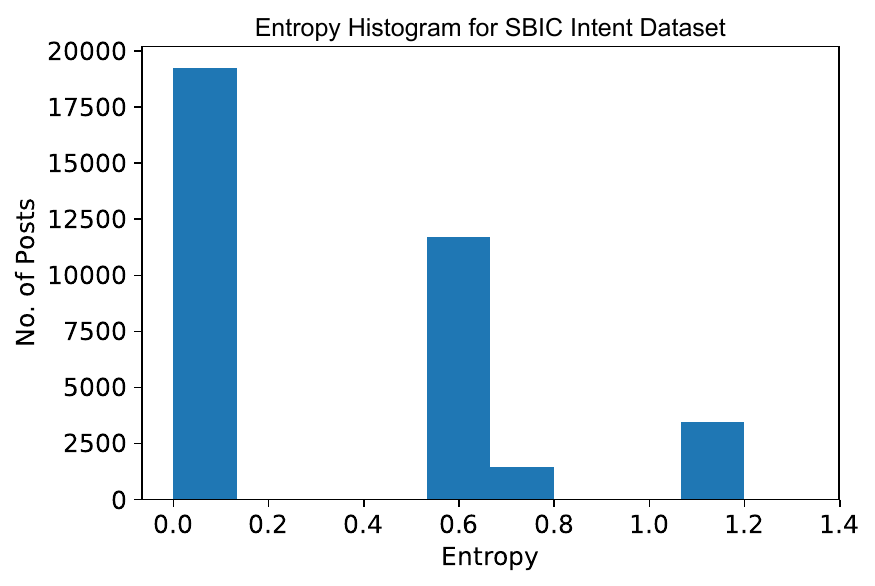}
\end{minipage}%
\hfill
\caption{Histograms of the label entropies per data item for each dataset. The histograms show similarities in the distributions of $\mathcal{D}_\texttt{FB}$ and JQ3, with a high relative level of entropy for the majority of their data items. On the other hand, the $\mathcal{D}_\texttt{JQ3}$ and $\mathcal{D}_\texttt{GE}$ datasets both have relatively large label spaces and both allow annotators to provide more than label per annotator per item, yet in terms of entropy distributions they are not similar. See Table 2 (main paper) for an overall summary of the datasets. \label{fig:entropy_graphs}}
\end{figure*}

See Figure~\ref{fig:entropy_graphs} for the Histograms.
\section{Model Selection Parameters}
\begin{table}[H]
\resizebox{\columnwidth}{!}{
  \begin{tabular}{c|c|ccccc}
          \textbf{Dataset} &          & $w = 0$    & $w=0.25$    & $w=0.50$    & $w=0.75$  & $ w=1 $      \\
          \hline
  \multicolumn{7}{c}{\textbf{Neighborhood Based Pooling Model}}                             \\\hline
  $\mathcal{D}_\texttt{FB}$      & $r$      & 0.8 & 1.4 & \cellcolor{blue!25} 3.0 & 3.6 & 4.6  \\
          & KL &  0.085 & 0.093 &  \cellcolor{blue!25}0.070 & 0.080 & 0.098 \\ \hline
$\mathcal{D}_\texttt{GE}$ & $r$ &  \cellcolor{blue!25}0.8 & 1.1 & 0.6 & 0.9 & 10.6 \\
    & KL  &   \cellcolor{blue!25}0.020 & 0.032 & 0.252 & 0.363 &  0.232 \\ \hline
  $\mathcal{D}_\texttt{JQ1}$     & $r$      & 3.5     & \cellcolor{blue!25} 5.6   & 3.4   & 5.6 & 2.8    \\
          & KL & 0.133 & 0.123 \cellcolor{blue!25} & 0.120 & 0.131 & 0.456  \\ \hline
  $\mathcal{D}_\texttt{JQ2}$     & $r$      & 3.2     & 3.5   & 2.4   & \cellcolor{blue!25} 2.8  & 5.5    \\
          & KL & 0.134 & 0.135 & 0.137 & \cellcolor{blue!25} 0.133 & 0.512 \ \\ \hline
  $\mathcal{D}_\texttt{JQ3}$     & $r$      &  \cellcolor{blue!25}10.2     & 5   & 6.1  & 8.7    &  3\\
          & KL & \cellcolor{blue!25} 0.023 & 0.024 & 0.027 & 0.028 & 0.884  \\ \hline
$\mathcal{D}_\texttt{SI}$ & $r$ &  \cellcolor{blue!25}2.4 & 9.3 & 4.8 & 9.8 & 11.4 \tabularnewline
    & KL &   \cellcolor{blue!25}0.160 & 0.176 & 0.180 & 0.190 &  0.350 \tabularnewline
\hline 
        \end{tabular}
        }
\caption{We achieve optimal label aggregation models on each label set with the presented neighborhood sizes ($r$) and KL-divergence (KL) for the datasets using NBP with KL-divergence as the loss function.}
\label{table:model_selection_nbp}
\end{table}
\begin{table*}[h!]
\resizebox{\textwidth}{!}{%
\begin{tabular}{c|c|ccccc||ccccc}
\textbf{Dataset}  &  & $w=0$  & $w=0.25$  & $w=0.50$  & $w=0.75$  & $w=1$  & $w=0$  & $w=0.25$  & $w=0.50$  & $w=0.75$  & $w=1$ \tabularnewline
\hline 
 &  & \multicolumn{5}{c||}{\textbf{FMM Model}} & \multicolumn{5}{c}{\textbf{GMM Model}}\tabularnewline
\hline 
$\mathcal{D}_\texttt{FB}$  & $p$  & \cellcolor{blue!25}4 & 30 & 36 & 4 & 32  & 26 & \cellcolor{blue!25}17 & 37 & 26 & 11 \tabularnewline
 & KL  &  \cellcolor{blue!25}0.704 &  1.551 &  1.587 & 1.273 &  1.598   &  0.702 &  \cellcolor{blue!25}0.696 &  0.706 &  0.702 &  1.432 \tabularnewline \hline

$\mathcal{D}_\texttt{GE}$ & $p$ & \cellcolor{blue!25}24 & 36 & 6 & 16 & 20 & \cellcolor{blue!25}25 & 34 & 24 & 26 & 26 \tabularnewline
   & KL &  \cellcolor{blue!25}2.053 &  2.121 & 3.312 &  3.941 &  4.804 & \cellcolor{blue!25}2.191 &  2.361 &  3.460 &  3.442 &  5.198 \tabularnewline \hline
$\mathcal{D}_\texttt{JQ1}$  & $p$  & 15 & \cellcolor{blue!25}6 & 7 & 9 & 6 & \cellcolor{blue!25}31 & 11 & 36 & 27 & 4 \tabularnewline
 & KL  & 0.465 &  \cellcolor{blue!25}0.458 &  0.468 &  0.461 & 0.903  & \cellcolor{blue!25}0.497 & 0.714 & 0.770 & 0.785 & 0.751 \tabularnewline
\hline 
$\mathcal{D}_\texttt{JQ2}$  & $p$  & 9 & \cellcolor{blue!25}8 & 5 & 5 & 5 & \cellcolor{blue!25}34 & 14 & 30 & 23 & 6  \tabularnewline
 & KL  &0.516 &  \cellcolor{blue!25}0.511 &  0.514 &  0.514 & 1.194  &  \cellcolor{blue!25}0.537 & 0.826 & 0.876 & 0.869 & 0.878 \tabularnewline
\hline 
$\mathcal{D}_\texttt{JQ3}$  & $p$  &  \cellcolor{blue!25}9 & 20 & 8 & 21 & 10  & 17 & \cellcolor{blue!25}24 & 37 & 23 & 11  \tabularnewline
 & KL  & \cellcolor{blue!25}0.965 &  1.406 & 1.371 &  1.586 &  1.457 & 0.903 &  \cellcolor{blue!25}0.902 &  0.918 &  0.905 &  1.491 \tabularnewline \hline 
$\mathcal{D}_\texttt{SI}$ & $p$ & 21 & 30 & 37 & 4 & \cellcolor{blue!25}5 & 12 & \cellcolor{blue!25}13 & 10 & 35 & 33  \tabularnewline
    & KL & 0.942 &  0.940 &  0.932 & 0.566 & \cellcolor{blue!25}0.355 & 0.849 &  \cellcolor{blue!25}0.711 &  1.935 &  1.989 & 1.932  \tabularnewline
\hline 
 &  & \multicolumn{5}{c||}{\textbf{K-Means Model}} & \multicolumn{5}{c}{\textbf{LDA Model}}\tabularnewline
\hline 
$\mathcal{D}_\texttt{FB}$  & $p$  & \cellcolor{blue!25}21 & 35 & 34 & 30 & 32 &  9 & \cellcolor{blue!25}19 & 16 & 5 & 8 \tabularnewline
 & KL  &\cellcolor{blue!25}0.702 &  0.710 &  0.733 &  0.705 &  0.715  & 0.680 &  \cellcolor{blue!25}0.584 &  0.687 & 0.689 & 0.690\tabularnewline
\hline 
$\mathcal{D}_\texttt{GE}$ & $p$ & \cellcolor{blue!25}27 & 34 & 19 & 31 & 28 & \cellcolor{blue!25}14 & 17 & 14 & 4 & 17 \tabularnewline
     & KL & \cellcolor{blue!25}2.322 &  2.593 &  3.541 &  4.430 &  4.293 &  \cellcolor{blue!25}1.907 &  1.997 &  1.985 & 2.494 &  2.938 \tabularnewline \hline
$\mathcal{D}_\texttt{JQ1}$  & $p$  &35 & \cellcolor{blue!25}21 & 35 & 35 & \cellcolor{blue!25}22 &  37 & 35 & \cellcolor{blue!25}14 & 22 & 10 \tabularnewline
 & KL  &0.471 &  \cellcolor{blue!25}0.463 &  0.467 &  0.477 &  \cellcolor{blue!25}0.463 & 0.450 &  0.449 &  \cellcolor{blue!25}0.435 &  0.480 &  0.470 \tabularnewline
\hline 
$\mathcal{D}_\texttt{JQ2}$  & $p$  & 11 & \cellcolor{blue!25}16 & 34 & 30 & 33 & \cellcolor{blue!25}19 & 7 & 5 & 19 & 9  \tabularnewline
 & KL  &0.515 &  \cellcolor{blue!25}0.512 &  0.540 &  0.519 &  0.538 & \cellcolor{blue!25}0.500 & 0.510 & 0.512 &  0.509 & 0.514  \tabularnewline
\hline 
$\mathcal{D}_\texttt{JQ3}$  & $p$  & 35 & 19 & 29 & \cellcolor{blue!25}14 & 32 & 5 & 5 & 4 & \cellcolor{blue!25}5 & 18 \tabularnewline
 & KL  & 0.969 &  0.938 &  0.948 &  \cellcolor{blue!25}0.912 &  0.953  &  0.889 & 0.887 & 0.886 & \cellcolor{blue!25}0.880 &  0.890 \tabularnewline
\hline 
$\mathcal{D}_\texttt{SI}$ & $p$ &  38 & 19 & 17 & 31 & \cellcolor{blue!25}35 &  6 & 15 & 4 & 18 & \cellcolor{blue!25}31 \tabularnewline
    & KL & 0.856 &  0.564 &  0.108 &  0.100 & \cellcolor{blue!25}0.050 & 0.935 &  0.935 & 0.496 &  0.397 &  \cellcolor{blue!25}0.296  \tabularnewline
\hline 
        \end{tabular}
    }
\caption{We achieve optimal label aggregation models on each dataset with the presented number of clusters ($p$) and KL-divergence (KL) for the datasets using the cluster sampler with KL-divergence as the loss function. $N_{max}$ is the number of items out of training set (1,000 items) assigned to largest cluster for  optimal label aggregation model ($p$). The mixing parameter ($w$) varies between $[0,1]$, where $w=0$ is a special case of pooling with only labels \cite{Liu2019HCOMP,Weerasooriya2020}. \emph{The lowest KL-divergence per dataset is highlighted in blue.}}
\label{table:model_selection}
\end{table*}

\end{document}